\documentclass[11pt]{article}
\pdfoutput=1
\usepackage{jheppub}
\usepackage{graphicx}
\usepackage{amssymb}
\usepackage{amsmath,amssymb}
\usepackage{slashed}
\usepackage{hyperref}
\usepackage{caption}
\usepackage{xcolor}
\usepackage{dsfont,soul}
\usepackage{verbatim}
\usepackage{subfig}
\usepackage{amsmath,amssymb,amscd,amsfonts,mathtools,csquotes}
\usepackage{xcolor}
\definecolor{markgreen}{RGB}{230,243,230}
\definecolor{darkolivegreen}{rgb}{0.33, 0.42, 0.18}
\definecolor{darkpastelgreen}{rgb}{0.01, 0.75, 0.24}

\usepackage[toc,page]{appendix}
\usepackage{epsfig}
\usepackage{epstopdf}
\usepackage{latexsym}
\usepackage{graphicx}
\usepackage{booktabs}
\usepackage{bbm}
\usepackage{color}
\usepackage{physics}
\usepackage{tensor}
\usepackage{verbatim}
\usepackage{subfig}
\usepackage{tikz}
\usepackage{ifthen}
\usetikzlibrary{matrix}
\usetikzlibrary{decorations.markings,calc,shapes,decorations.pathmorphing,patterns,decorations.pathreplacing,arrows.meta}
\usetikzlibrary{positioning}
\usepackage{hyperref}
\usetikzlibrary{patterns}
\usetikzlibrary{positioning}
\tikzset{snake it/.style={decorate, decoration=snake}}

\pdfoutput=1
\makeatletter
\def\@fpheader{\relax}
\makeatother

\newboolean{showcomments}
\setboolean{showcomments}{false}
\newcommand\rem[1]{\ifthenelse{\boolean{showcomments}}{{#1}}{}}
\newcommand{\be}{\begin{equation}}
\newcommand{\ee}{\end{equation}}

\title{\Large The Fate of Information Localizability and Holography in Quantum Gravity}

\author{Hao Geng$^{a,b}$, Daniel Jafferis$^{b}$, Pushkal Shrivastava$^{b}$ and Neeraj Tata$^{b}$}
\affiliation{$^{a}$
Gravity, Spacetime, and Particle Physics (GRASP) Initiative, Harvard University, 17 Oxford St., Cambridge, MA, 02138, USA.}
\affiliation{$^{b}$ Jefferson Physical Laboratory, Harvard University, 17 Oxford St., Cambridge, MA, 02138, USA.}
\emailAdd{haogeng@fas.harvard.edu, jafferis@g.harvard.edu, pushkalshrivastava@g.harvard.edu, ntata@g.harvard.edu}

\abstract{The AdS/CFT correspondence states an equivalence between a quantum gravitational theory in a (d+1)-dimensional anti-de Sitter spacetime (AdS$_{d+1}$) and a d-dimensional conformal field theory (CFT$_{d}$). The CFT$_{d}$ lives on the asymptotic boundary of the bulk AdS$_{d+1}$. Hence  a local operator in the bulk of the AdS$_{d+1}$ should be reconstructable using operators living on the asymptotic boundary at the same instant. 
The existence of such a reconstruction is highly nontrivial and is conceptually puzzling if we think in terms of physically detecting a local bulk particle from the boundary of the AdS$_{d+1}$, as this signals a 
non-local information encoding scheme. In this paper, we explore situations where such non-locally encoded information can be observed in semiclassical gravity. We study examples where it is more efficient to utilize such effects in quantum gravity to detect a bulk excitation than to wait for signals to reach the boundary. Furthermore, we provide exemplified situations
for which the protocol fails, and the 
non-locality of information
is suppressed. These exemplified scenarios can be taken as explicit examples of the emergence of a perturbatively localized observer. In such cases, holography cannot be proven  at the perturbative level in Newton's constant $G_{N}$ via the non-localizability of information. }

\begin{document}
\maketitle
\flushbottom
\pagebreak
\section{Introduction}\label{sec:intro}
Recent progress in the study of the black hole information paradox suggests that the fact that information is non-locally encoded in quantum gravity plays a key role in the resolution of this paradox \cite{Almheiri:2019psf,Penington:2019npb,Geng:2020qvw,Geng:2021hlu,Geng:2023zhq}. This is not a surprise from the perspective of the AdS/CFT correspondence \cite{Maldacena:1997re,Gubser:1998bc,Witten:1998qj}, which conjectures an exact equivalence between a quantum gravitational theory in the bulk of AdS$_{d+1}$ and a conformal field theory CFT$_{d}$ that lives on the asymptotic boundary of this AdS$_{d+1}$. Hence all physical information in the bulk of a gravitating spacetime is encoded on its boundary, in a non-local manner. A typical example of such an effect is the gravitational Gauss' law. The question we address is how large the lack of locality is and in what sense it is physically measurable, in terms of the finiteness of the time and sensitivity required.\footnote{For the sake of convenience of discussion, we will refer to the above non-local effect as ``breakdown of locality" in this paper.}  For example, if the time it takes to measure such non-local effects is longer than the causal propagation time in the bulk or if the amplitude of such non-local effects is sufficiently suppressed compared to its variance in a given state, then they are not physical.

These questions were recently studied in the context of the AdS/CFT correspondence in \cite{Chowdhury:2020hse,Chowdhury:2021nxw} and in asymptotically flat space in \cite{Laddha:2020kvp,Chowdhury:2022wcv} (see \cite{Raju:2020smc,Raju:2021lwh} for reviews). In these interesting works, compelling nonperturbative arguments for the statement that the bulk information is always encoded on the asymptotic boundary are provided, proofs for this statement are given in perturbation theory around empty spacetime, and a physical protocol for a near-boundary observer to extract the bulk information with high precision is provided. These works suggest that bulk locality already breaks down perturbatively in the expansion in terms of the Newton's constant $G_{N}$ (i.e. low energy theory of gravity), 
which gives a perturbative version of holography. Nonetheless, it remains unclear whether such a breakdown of locality can be measured efficiently without access to multiple copies of the system. The reason is that the protocols in \cite{Chowdhury:2020hse,Chowdhury:2022wcv}  aimed to determine the full state from the boundary and 
therefore require the performance of an exponentially large number of measurements. Thus, operating such protocols requires the access to multiple copies of the system or a large amount of time.

At this stage, a natural question arises:
\begin{displayquote}
\textit{\textbf{Is it ever possible to utilize the effects of the breakdown of locality in quantum gravity to efficiently detect a bulk excitation from the boundary of spacetime using only a single copy of the system?}}
\end{displayquote}
Here, by efficient, we mean that the time taken to perform the measurements is less than the time it takes for the signal from the bulk to reach the boundary. We emphasize that, unlike the lofty goal of determining the full state \cite{Chowdhury:2020hse}, we are only concerned with the detection of a unitary bulk excitation. Such a bulk excitation is created by a unitary operator and would be purely localized in the bulk if there were no gravity. However, once gravity is turned on, and if the background state has isometries, then the excitation in question will not be localized in the bulk due to its Gauss law dressing. Thus, it is possible for the above question to have a positive answer.

\begin{figure}[h]
\begin{centering}
\includegraphics[width=0.25\textwidth]{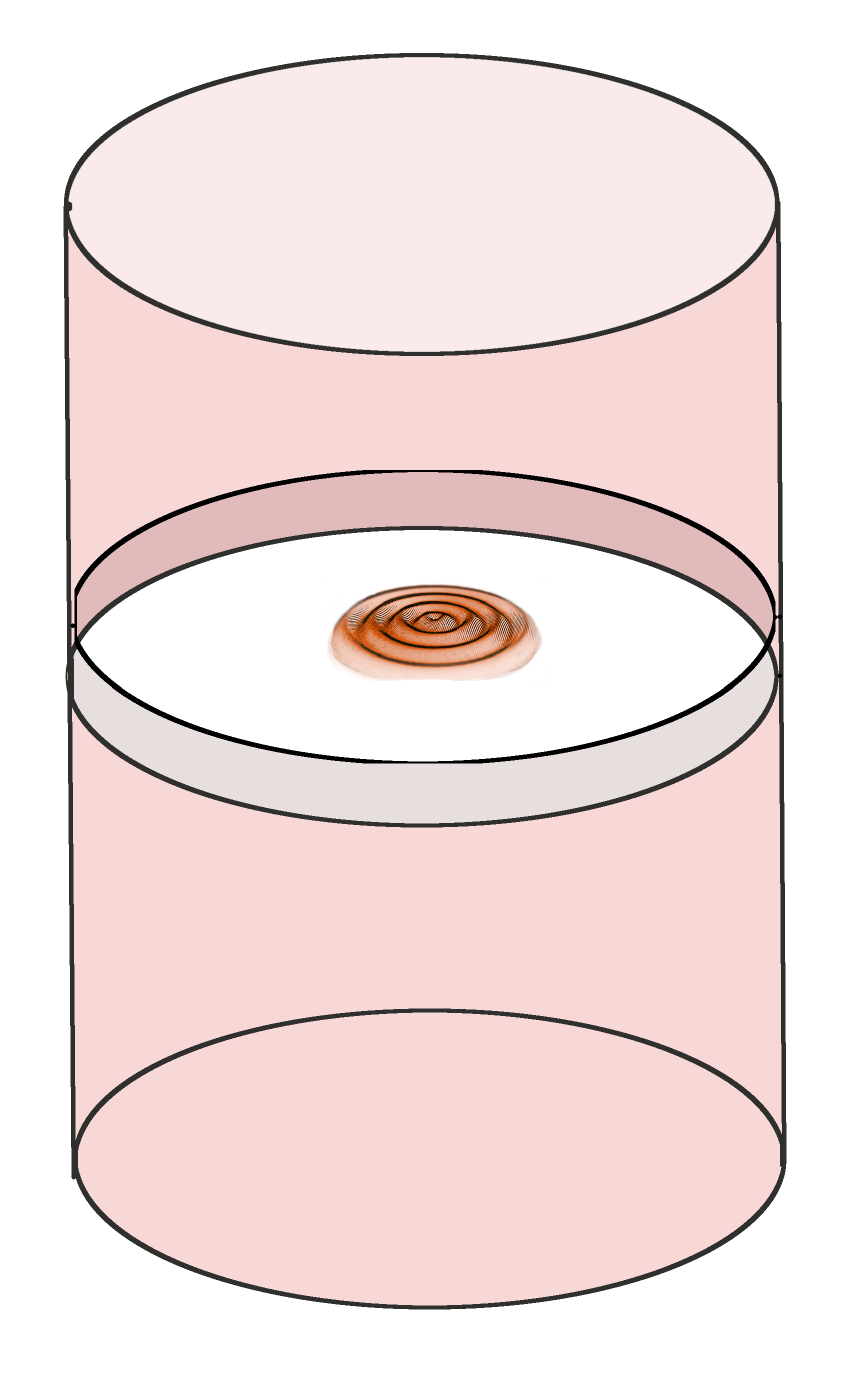}
\caption{\small{\textit{An illustration of the particle detection measurement in global AdS$_{d+1}$. The white slice is a Cauchy slice in the bulk of AdS$_{d+1}$ and the gray ring is near its asymptotic boundary. A particle excitation, the orange wavepacket, is located at the center of a bulk Cauchy slice and an observer near the boundary at the same Cauchy slice, for example inside the gray ring, is trying to detect this particle.}}}
\label{pic:demo}
\end{centering}
\end{figure}

The goal of the paper is to explore the above question in asymptotically AdS spacetimes. Based on \cite{Laddha:2020kvp}, we provide examples in the AdS/CFT context where we can make use of the aforementioned perturbative breakdown of locality to efficiently detect local bulk particles in an empty background from the boundary (see Fig.\ref{pic:demo}). This measurement protocol is efficient as it only requires measurements of simple boundary operators\footnote{Here the boundary operators really mean the boundary CFT operators $\hat{\mathcal{O}}(x)$ which can be thought of as extrapolated versions of the near-boundary bulk operators $\hat{\phi}(z,x)$ as $\hat{\mathcal{O}}(x)=\lim_{z\rightarrow0}z^{-\Delta}\hat{\phi}(z,x)$ according to the standard rules in AdS/CFT \cite{Gubser:1998bc,Witten:1998qj}. As will be discussed in Sec.~\ref{sec:giddingsexp}, this avoids the issue of ``exponential smallness" of the response of the boundary operators to local bulk excitations discussed in \cite{Giddings:2021khn}.} for a finite number of times and we can also show that each measurement can be carried out in a fairly short time. This is due to the simplicity of the task which though demonstrates that the perturbative breakdown of locality is physical at least in the setup we will consider, as we will explain later. Moreover, this protocol requires a small but nonzero $G_{N}$ which signals its gravitational nature and suggests its potential importance in quantum gravity. 

We note that another way to measure the bulk particles from the boundary is to simply wait for the particles to propagate to the boundary and collect them. Let's denote the time it takes for this propagation to be $t_{d}$. For the aforementioned protocol, leveraging the perturbative breakdown of locality, to work efficiently, one wants to finish such measurements before the particles arrive at the boundary. In fact, the simple boundary measurements that we perform are the correlators between the Hamiltonian $\hat{H}$ and simple local operators $\hat{\mathcal{O}}(x)$ \cite{Chowdhury:2020hse}. However, an interesting effect we discover in this paper is that such a measurement will backreact on the relative time shift between the bulk and the boundary. Hence, if this time shift is too big, say larger than $t_d$, then the particle will already be at the boundary, while such a measurement is being performed. This would render such a protocol less efficient. Interestingly, we find that the backreaction on this relative time shift depends on the precision of the measurement, the higher the precision the stronger the backreaction. This backreaction is a gravitational effect. Hence, we conclude that such measurement cannot be arbitrarily precise if one doesn't want to destroy the background by causing a huge gravitational backreaction.

Next we try to address the question if this perturbative breakdown of locality is of any relevance in more generic situations. We first study this question by considering the situation that the bulk now has a nontrivial background configuration of matter distributions $\phi^{0}(x)$. In this case, we are able to construct a local bulk operator that satisfies the diffeomorphism constrains by dressing it to this background configuration. We can show that the protocol fails to detect the bulk excitation created by such an operator. Nevertheless, we will see that this operator is not a well-defined bulk operator if the background configuration is not strong enough.\footnote{This is in analogy with the electromagnetism case that if we want to dress a local charged operator to some background charge distribution then the charge of the background distribution has to be larger than the charge of the to be dressed operator (and with an opposite sign).} Hence in the limit where the background configuration is weak we recover the situation where our protocol could work to detect bulk particles. In such situations, it will appear that classically there are undetectable localized bulk operators; however, quantumly our protocol probabilistically succeeds. This provides a one-parameter family of constructions of bulk operators that interpolates between the situations where the perturbative version of holography works and the situation where it breaks down. 

This construction improves on those in \cite{Bahiru:2022oas,Bahiru:2023zlc} by providing an explicit construction of the operator in the bulk theory with a clear bulk interpretation, and is fully quantum as opposed to \cite{Folkestad:2023cze}. More importantly, this construction suggests a perturbative tension between locality and the existence of simple boundary protocols to detect the bulk information as we couldn't have both of them at the perturbative level. This further suggests that practical usage of holography in such a context with a nontrivial background matter distribution, which is not yet a black hole, deserves a better understanding. Second, we construct situations where we could have an approximate notion of locality in the bulk. We emphasize that we work to the leading order of nonvanishing $G_{N}$ (where we have set the AdS length length scale to one) and our results can be easily extended to all orders in perturbation theory following \cite{Chowdhury:2021nxw}. Then, we extend the above constructions to novel situations where we can dress operators to intrinsically quantum features of the state. Interestingly, in these situations, one can interpret the operators as to be dressed to an observer. We provide two types of constructions of this kind: 
\begin{enumerate}
    \item We construct an emergent observer, i.e. a clock, out of the system. We make use of the level spacing of the matter energy spectrum. The clock becomes perfect as the entropy of the background state goes to infinity, i.e. it is a linear combination of densely distributed energy eigenstates.
    \item We consider the situations with an explicit observer existing outside the system we are considering. The observer is entangled with the AdS system. We show that we can dress operators inside the AdS to the external observer using this entanglement. An intuitive picture is that the dressed operators are in fact connected to the external observer by \textit{quantum wormholes} \cite{Geng:2025rov,Geng:2025byh}.
\end{enumerate}
These constructions are the technical novelties in our work. The second construction also resolves a point recently raised in \cite{Antonini:2025sur}.

Finally, we study the perturbative localizability of information in the bulk from the dual CFT point of view. From the CFT perspective, whether one could have perturbatively localized operators in the bulk depends on whether there exist a certain type of operators that correlate with the stress-energy tensor. These operators can be interpreted as the CFT dual of the observer. Hence, the results are consistent with the situations in the bulk.

This paper is organized as follows. In Sec.~\ref{sec:no locality} we set up the question we are studying, provide the bulk particle detection protocol and show why it works practically in detail. In Sec.~\ref{sec:no holography} we consider situations where the particle detection protocol in Sec.~\ref{sec:no locality} fails by constructing local bulk operators which couldn't be perturbatively detected by boundary observers and we show that this requires a strong enough background configuration, for example a classical matter distribution. In addition, we construct situations where we could have an approximate notion of locality in the bulk in Sec.~\ref{sec:no holography}. In Sec.~\ref{sec:CFT} we provide the dual CFT perspective on the perturbative localizablity of information in the bulk. In Sec.~\ref{sec:relation} we discuss the relation between our work and the previous work in the literature and point out some future questions to study. In Sec.~\ref{sec:con} we conclude our paper with discussions.

\section{An Efficient Detection of Bulk Particles Using Simple Boundary Operators in AdS/CFT}\label{sec:no locality}
As mentioned in the introduction, the goal of this paper is to explore the fate of locality in quantum gravity. More precisely, we explore the possibility of obtaining information via measurements exploiting the non-local nature of the constraint equations in quantum gravity. It is expected that quantum gravity stores information holographically \cite{Raju:2019qjq, Laddha:2020kvp, Chowdhury:2020hse, Chowdhury:2021nxw, Raju:2021lwh, Chakraborty:2023los} (see also \cite{Marolf:2006bk,Giddings:2006vu,Marolf:2008mf,Marolf:2008tx,Donnelly:2015hta,Donnelly:2016auv,Donnelly:2016rvo,Donnelly:2017jcd} for some relevant earlier work). Hence, in principle, it is possible to determine the quantum state just from the boundary of a bulk Cauchy slice. This leads to an interesting feature of quantum gravity that is precluded in a local quantum field theory. 

Suppose we foliate the spacetime by Cauchy slices labeled by a boundary time $t$. Now, consider a unitary excitation created in the bulk at a fixed boundary time, say $t=0$. In a local theory with no gravity, such an excitation can be detected near the boundary only after a signal from the excitation propagates to and arrives at the boundary. This signal can propagate at most at the speed of light. Hence, there would be a time delay between the \emph{occurrence} of the excitation in the bulk and its \emph{detection} near the boundary. Let us denote this time delay by $t_d$. Hence, in the cases without gravity, the information is localizable. 

In quantum gravity, since the state can be determined from the boundary of the bulk Cauchy slice, one may expect that there is no time delay between the \emph{occurrence} and \emph{detection} of the excitation. However, it is unclear whether such measurements can be performed in semiclassical gravity and in a time that is shorter than the previously discussed time delay, $t_d$. 

We will see that such a measurement will cause a backreaction on the relative time shift between boundary and the bulk. This backreaction would be large if the measurement precision is high. Thus, if the measurement is too precise, such that this backreacted time shift is larger than $t_{d}$, the particle will already be at the boundary while the measurement is being performed. However, we will see that this backreaction can be extremely suppressed if we only want to perform coarse-grained measurements, such as detecting the existence of a bulk particle from the boundary without knowing too much about the details.

For the purpose of this section, we restrict to quantum gravity in empty anti-de-Sitter space. First, we review the aspects of quantum field theory in anti-de-Sitter space that would be relevant to our discussions.

\subsection{Quantum Field in AdS}
The metric of a $D=d+1$ dimensional anti-de-Sitter space in global coordinates (in AdS units, i.e., $\ell_{AdS} = 1$) is:
\begin{equation}
	ds^2 = {-dt^2 + d\rho^2 + \sin^2\rho \, d\Omega_{d-1}^2 \over \cos^2 \rho}\,,\label{eq:background}
\end{equation}
The coordinate $\rho$ ranges from $0$ at the center of AdS to ${\pi\over 2}$ at its asymptotic boundary.
A free real massive scalar field satisfies the following equation of motion
\begin{equation}
	(\Box - m^2 ) \phi = 0\,.
\end{equation}
We can expand the quantum scalar field using the spherical harmonic basis.
\begin{equation}
	\begin{split}
	\hat{\phi}(\rho, t, \Omega) = \sum\limits_{n , \vec{\ell}} a_{n, , \vec{\ell}} \, e^{- i \omega_{n,\ell} t} \psi_{n,\ell}(\rho) \, Y_{\vec{\ell}}(\Omega) + h.c. \,.
	\end{split}
\end{equation}
In the above equation, we denote the spherical harmonic modes on $S^{d-1}$ by $\vec{\ell} = (\ell, m, \ldots) $. The energy of the modes is quantized as follows
\begin{equation}
	\begin{split}
	\omega_{n,\ell} = \Delta + 2n + \ell ,\qquad  \Delta = {d\over2} + \sqrt{{\left({d\over 2}\right)^2 + m^2} }\,,
	\end{split}
\end{equation}
for which we are doing the standard quantization \cite{Balasubramanian:1998sn} in which energy quantization  comes from imposing regularity conditions on the field $\hat{\phi}(\rho,t,\Omega)$ at $\rho=0$, with $n=0,1,2,\cdots$.
The coefficients $a_{n,\vec{l}}$ and $a^{\dagger}_{n,\vec{l}}$ of the modes satisfy the commutation relation of creation and annihilation operators in canonical quantization. These commutators are canonically normalized as
\begin{equation}
	[a_{n , \vec{\ell}} , a_{n' , \vec{\ell\,}'}^{\dagger} ] = \delta_{n,n'}\, \delta_{\vec{\ell\,} , \vec{\ell\,}'}\,,
\end{equation}
which requires
\begin{equation}
	\begin{split}
	\psi_{n,\ell}(\rho) &= {1\over N_{n,\ell}} \sin^{\ell} \rho \, \cos^{\Delta}\rho \, \ _2F_1 (-n,\Delta + \ell + n, \ell +{d\over 2} , \sin^2\rho)\,, \\ 
	 N_{n,\ell} &= (-1)^n \sqrt{\Gamma(n+1) \Gamma^2(\ell+{d\over 2}) \Gamma(\Delta + n - {d-2\over2}) \over \Gamma(n+\ell + {d\over 2} ) \Gamma(\Delta+n+\ell) }\,.
	\end{split}
\end{equation}

\subsection{Problem statement}
We want to ask the question that if a sufficiently localized bulk excitation when gravity is decoupled can be detected by a boundary observer or not in the geometry \eqref{eq:background} when gravity is dynamical, i.e. $G_{N}$ is not exactly zero. We create such an excitation by a unitary operator $U$ constructed from a smeared local bulk field operator $\hat{\phi}$ using a sufficiently locally supported kernel $f(t,\rho)$ as\footnote{This setup was explained in \cite{Laddha:2020kvp}.}
\begin{equation}
U=e^{i\alpha\int d\rho d\Omega dt f(\rho,t)\hat{\phi}(\rho,t,\Omega)}\,,\label{eq:unitary}
\end{equation}
and the excited state is given by 
\begin{equation}
\ket{\Psi}=U\ket{0}\,,
\end{equation}
where $\ket{0}$ is the ground state of the free massive real scalar $\hat{\phi}(t,\rho,\Omega)$ in the background geometry Equ.~(\ref{eq:background}). A boundary observer is performing measurements of boundary operators and in semiclassical gravity the boundary observer is performing the measurements of simple boundary operators and their low order correlators. In our context, the set of simple boundary operators consists of finite order polynomials of the boundary extrapolation of bulk fields, for example 
\begin{equation}
\hat{O}(t,\Omega)=\lim_{\rho\rightarrow\frac{\pi}{2}}\cos^{-\Delta}\rho\hat{\phi}(t,\rho,\Omega)\,,\label{eq:O}
\end{equation}
and the energy or the ADM Hamiltonian operator
\begin{equation}
\hat{H} =\frac{1}{\sqrt{16\pi G_{N}}} \lim_{\rho\rightarrow\frac{\pi}{2}}\cos^{-d+2}\rho \int
\Big(\nabla_{i}\hat{h}^{i\rho}(\rho, t, \Omega)-\nabla^{\rho}\hat{h}(\rho, t, \Omega)\Big)d\Omega\,,\label{eq:H}
\end{equation}
where $\hat{h}_{\mu\nu}(t,\rho,\Omega)$ is the bulk graviton field whose kinetic term is canonically normalized and we take the ADM convention \cite{Arnowitt:1962hi} (to be discussed in detail in Sec.~\ref{sec:no holography}).

\subsection{QFT in Fixed Background Doesn't Work}
In the case that we consider a quantum field theory in a fixed background Equ.~(\ref{eq:background}), gravity is not dynamical so there is no graviton field.\footnote{More precisely to measure the energy the boundary observer is in fact measuring $\hat{H}_{\partial}=\sqrt{16\pi G_{N}}\hat{H}$ where $\hat{H}$ is the ADM Hamiltonian $\hat{H}$ in Equ.~(\ref{eq:H}). This operator $\hat{H}_{\partial}$ is zero in the decoupling limit $G_{N}\rightarrow0$ so in this decoupling limit the boundary observer doesn't have the access to the Hamiltonian. } The basic property of quantum field theory is \textit{locality}. This ensures that measurements performed by a boundary observer on the same Cauchy slice as the excitation $U$ cannot distinguish $\ket{\Psi}$ from $\ket{0}$. This is because
\begin{equation}
\bra{\Psi}\hat{O}(t,\Omega)\ket{\Psi}=\bra{0}U^{\dagger}\hat{O}(t,\Omega)U\ket{0}=\bra{0}\hat{O}(t,\Omega)U^{\dagger}U\ket{0}=\bra{0}\hat{O}(t,\Omega)\ket{0}\,,
\end{equation}
where in the second step we use locality and the fact that the operators $U$ and $\hat{O}(t,\Omega)$ are supported on spacelikely separated regions and in the last step we used the unitarity of the operator $U$. This result can be easily extended to correlators of any boundary operators. Hence locality in quantum field theory ensures that the aforementioned task cannot be finished by a near-boundary observer.

\subsection{Protocol with gravity}
Now we consider the case where gravity is dynamical and meanwhile $G_{N}$ is small but nonzero. In this case, we can consider perturbation theory in $G_{N}$ around the background Equ.~(\ref{eq:background}) as a good approximation.

Consider the state
\begin{equation}
	|\Psi \rangle = U\ket{0}=e^{i \alpha \int d\rho d\Omega dt \,  f(\rho,t) \hat{\phi}(\rho,t,\Omega) } | 0 \rangle\,,\label{eq:psi}
\end{equation}
where $|0\rangle$ is the vacuum state and $f$ is a smearing function describing the profile of the created particle state. For example, if $f(\rho,t)$ is supported on a subset $\mathcal{B}$ of the bulk AdS$_{d+1}$ then the state $\ket{\Psi}$ can be thought of as containing an excitation localized in $\mathcal{B}$. Now the goal of the near-boundary observer is to detect this state, i.e. to distinguish that it is different from the ground state $\ket{0}$, and potentially reconstruct the profile $f(\rho,t)$. The essential difference between the current situation and the situation where gravity is not dynamical is that the near-boundary observer is now equipped with more operators which comes from the boundary extrapolation of the bulk graviton field $\hat{h}_{\mu\nu}$. An example of such an operator is the ADM Hamiltonian $\hat{H}$ Equ.~(\ref{eq:H}) which is equal to the Hamiltonian of the bulk matter field $\hat{\phi}(t,\rho,\Omega)$.\footnote{This result comes from the Hamiltonian constraint as will be discussed in detail in Sec.\ref{sec:no holography}. For the sake of the understanding of this section, this result can be taken heuristically as a basic statement of the AdS/CFT correspondence that the bulk and boundary Hamiltonian are the same.}

Following \cite{Laddha:2020kvp}, the explicit protocol for the near-boundary observer to finish the above task is to measure the following correlator
\begin{equation}
    \bra{\Psi}\hat{H}\hat{O}(t,\Omega)\ket{\Psi}\,,
\end{equation}
where $\hat{O}(t,\Omega)$ is the boundary extrapolation of the bulk field $\hat{\phi}(t,\rho,\Omega)$ following Equ.~(\ref{eq:O}). This correlator is clearly zero for ground state $\ket{0}$. For the state $\ket{\Psi}$ defined in Equ.~(\ref{eq:psi}), we will have
\begin{equation}
\begin{split}
\bra{\Psi}\hat{H}\hat{O}(t,\Omega)\ket{\Psi}&=\bra{0}U^{\dagger}\hat{H}\hat{O}(t,\Omega)U\ket{0}\,,\\&=\bra{0}U^{\dagger}\hat{H}U\hat{O}(t,\Omega)\ket{0}\,,\\&=\alpha\int d\rho'd\Omega'dt'f(\rho',t')\bra{0}\partial_{t'}\hat{\phi}(\rho',t',\Omega)\hat{O}(t,\Omega)\ket{0} \,,\\&=\alpha\int d\rho'd\Omega'dt' f(\rho',t')\partial_{t'}K_{\Delta}(\rho',t',\Omega'|t,\Omega) \,,\label{eq:measure}
\end{split}
\end{equation}
where in the second step we used the fact that the supports of the operators $U$ and $\hat{O}(t,\Omega)$ are spacelikely separated, in the third step we used the fact that $\hat{H}$ as the Hamiltonian generates the time evolution of the field operator $\hat{\phi}(\rho,t,\Omega)$, the function $K_{\Delta}(\rho',t',\Omega'|t,\Omega)$ is the bulk-to-boundary propagator in AdS/CFT \cite{Balasubramanian:1998sn} and we denoted $d\Omega=\sin^{d-1}\rho d\Omega_{d-1}$. Now with the measurement Equ.~(\ref{eq:measure}) and the knowledge of the bulk-to-boundary propagator $K_{\Delta}(\rho',t',\Omega'|t,\Omega)$ the near-boundary observer may simply reconstruct $f(\rho,t)$ by an involution with respect to the inverse propagator $K_{\Delta}^{-1}(\rho',t',\Omega'|t,\Omega)$. This inverse propagator satisfies
\begin{equation}
\int dtd\Omega K_{\Delta}(\rho',t',\Omega'|t,\Omega)K_{\Delta}^{-1}(\rho'',t'',\Omega''|t,\Omega)=\delta(\rho'-\rho'')\delta(t'-t'')\delta^{d-1}(\Omega'-\Omega'')\,.\label{eq:recon}
\end{equation}
We notice that the above protocol works perfectly for a simple question i.e. to detect the particle state or excitation described by the state Equ.~(\ref{eq:psi}) by distinguishing it from the ground state $\ket{0}$ as this only requires few measurements to see that we get a nonzero result for Equ.~(\ref{eq:measure}). 

We want to emphasize that, in fact, it is even possible to determine the profile $f(\rho,t)$ using Equ.~(\ref{eq:recon}). As shown in \cite{Laddha:2020kvp}, the kernel appearing in the correlator \eqref{eq:measure} is analytic in the lower half of complex-$t$ plane. Hence, from the measurements of the correlator in a neighborhood of a time band, it is possible to reconstruct the function $f$. However, this reconstruction requires a very precise measurement of the correlation function, and we will see that it takes a long time due to the uncertainty principle. Another similar way to reconstruct the profile from the precise result of Equ.~(\ref{eq:measure}) is to use the completeness relation of the bulk-boundary propagators \eqref{eq:recon}, which requires measurements over light-crossing time to know the explicit form of the time-dependence. The measurement involved is to determine the correlator between $\hat{O}(t,\Omega)$ and the Hamiltonian $\hat{H}$. We notice that it is not simply a measure of the energy in a state.

\subsection{Practical Relevance of the Protocol and Its Limitation}\label{sec:smearing}
The protocol to detect the excitation which is spacelike separated from the asymptotic boundary on a bulk Cauchy slice is practically relevant or physical only if it can be carried out in a finite time with a nontrivial result. Hence, we want to understand how long it takes for the near-boundary observer to measure a non-zero value for the correlator given in  Equ.~(\ref{eq:measure}), or more precisely, to know what the probability for the measurement result to be nonzero at each time of the measurement is. We will see that these questions are in fact subtle in the current context due to potential gravitational backreactions.

To address the above questions, we consider the following setup for the measurement of an observable $\hat{H}\hat{A}$, where $\hat{H}$ is the Hamiltonian and $\hat{A}$ is a boundary operator spacelikely separated from the bulk excitation. The operator $\hat{A}$ is understood to be supported only over a short time band in the asymptotic boundary of the AdS$_{d+1}$. Let us consider a pointer system with the position operator $\hat{X}$ and the conjugate momentum operator $\hat{P}$. We couple the AdS$_{d+1}$ system with the pointer system via a term $\delta \hat{H}_{\text{tot}}=\hat{H}\hat{A}(t)\hat{P}f_{c}(t)$, where the function $f_{c}(t)$ is supported over a short time band and can be taken as a delta-function, in the Hamiltonian of the combined system which consists of the pointer and the original AdS$_{d+1}$ system.

We work in the interaction picture and let $|\psi_i\rangle$ and $|\psi_f\rangle$ denote the initial and final states of the combined system before and after we have coupled the pointer and the original AdS$_{d+1}$ system. The initial state is written as 
\begin{equation}
\ket{\psi_{i}}=\ket{\Psi}\otimes\ket{X=0,\sigma_{X}}\,,
\end{equation}
and the final state can be written as
\begin{equation}
    |\psi_f \rangle = e^{-i  \hat{H}\hat{A}\hat{P}} |\psi_i\rangle
\end{equation}
The coupling has two effects:
\begin{enumerate}
    \item The coupling shifts the location $\hat{X}$ of the pointer;
    \item The coupling introduces a time evolution in the boundary of the AdS$_{d+1}$ relative to the bulk.
\end{enumerate}
As we will see, this time evolution in the second effect is a quantum effect and it can be understood as a gravitational backreaction.

We note that if the AdS$_{d+1}$ system is in the vacuum state, i.e. if $\ket{\Psi}=\ket{0}$, the operator $\hat{H}\hat{A}$ annihilates the state. Hence the coupling is immaterial and the final state $\ket{\psi_{f}}$ is same as the initial state $\ket{\psi_{i}}$. As the state of the pointer system doesn't change, we won't detect any excitation according to our particle detecting protocol.

However, if the initial state contains an excitation in the bulk, i.e. $\ket{\Psi}\neq\ket{0}$, the final state $\ket{\psi_{f}}$ would be different and this can potentially lead to a successful detection of the excitation contained in $\ket{\Psi}$. We shall consider it as a non-local detection if we are able to detect the excitation before it propagates to the boundary, \textit{i.e.} the relative time evolution introduced in between the boundary and the bulk is small enough.

The change in the pointer location is given by 
\begin{equation}
\begin{split}
    \Delta X &= \langle \psi_f| \hat{X} | \psi_f \rangle - \langle \psi_i| \hat{X} | \psi_i \rangle \\
    & = \langle \psi_i | \hat{X} +  \hat{H}\hat{A} |\psi_i \rangle  - \langle \psi_i| \hat{X} | \psi_i \rangle\\
    & = \langle \psi_i | \hat{H}\hat{A} | \psi_i \rangle = \langle\Psi|\hat{H}\hat{A}|\Psi\rangle\,.\label{eq:changeX}
\end{split}
\end{equation}

Similarly, the expected shift in the AdS$_{d+1}$ asymptotic time is given by 
\begin{equation}
   \langle\delta t_{AdS}\rangle=\langle \psi_i |\hat{A} \hat{P}| \psi_i \rangle=\bra{\Psi}\hat{A}\ket{\Psi}\bra{X=0,\sigma_{X}}\hat{P}\ket{X=0,\sigma_{X}}\,.
\end{equation}
This is zero as $\hat{A}$ is space-like separated from the unitary excitation in the bulk leading to $\langle\Psi|\hat{A}|\Psi\rangle$ = 0. However, we also have to make sure that the variance of this time shift is small. 
This variance can be computed as
\begin{equation}
\begin{split}
    \sigma_{\delta t_{AdS}} &= \sigma_A \, \sigma_P\,,\label{eq:sigmadt}
\end{split}
\end{equation}
where $\sigma_{(.)}$ denotes the fluctuation of the operator $(.)$ in the initial state $\ket{\psi_{i}}$. If the variance of this time shift is large then there is a large probability that the signal from the bulk excitation arrives at the boundary at a much earlier time. Thus, it might be cheaper in this case to measure the bulk excitation by waiting for its arrival at the boundary than reading our apparatus. Hence, we focus on the situations that the variance of this time shift is small comparing to the original light crossing time to fully take advantage of our protocol and we will comment on when this time shift is unsuppressed at the end.

Let us consider the pointer to be in a Gaussian initial state with the position distribution centered at $0$ and variance $\sigma_X$. From the uncertainty principle, the variance in momentum is $\sigma_P \sim  {1\over \sigma_X}$. The physical interpretation of the variance $\sigma_X$ is the resolution of the measuring apparatus, i.e. the pointer. For a successful detection, we should choose the resolution to be smaller than the pointer shift $\Delta X$  in Equ.~(\ref{eq:changeX}), resulting in a signal to noise ratio greater than 1. Motivated by this, we impose the detectability criterion:
\begin{equation}
    \sigma_P \ge {1\over |\langle\Psi|\hat{H}\hat{A}|\Psi\rangle|} \quad \implies \quad \sigma_{\delta t_{AdS}} \ge \abs{{\sigma_A \over\langle\Psi|\hat{H}\hat{A}|\Psi\rangle }}
\end{equation}

\begin{figure}[h]
    \centering
    \includegraphics[width=0.7\textwidth]{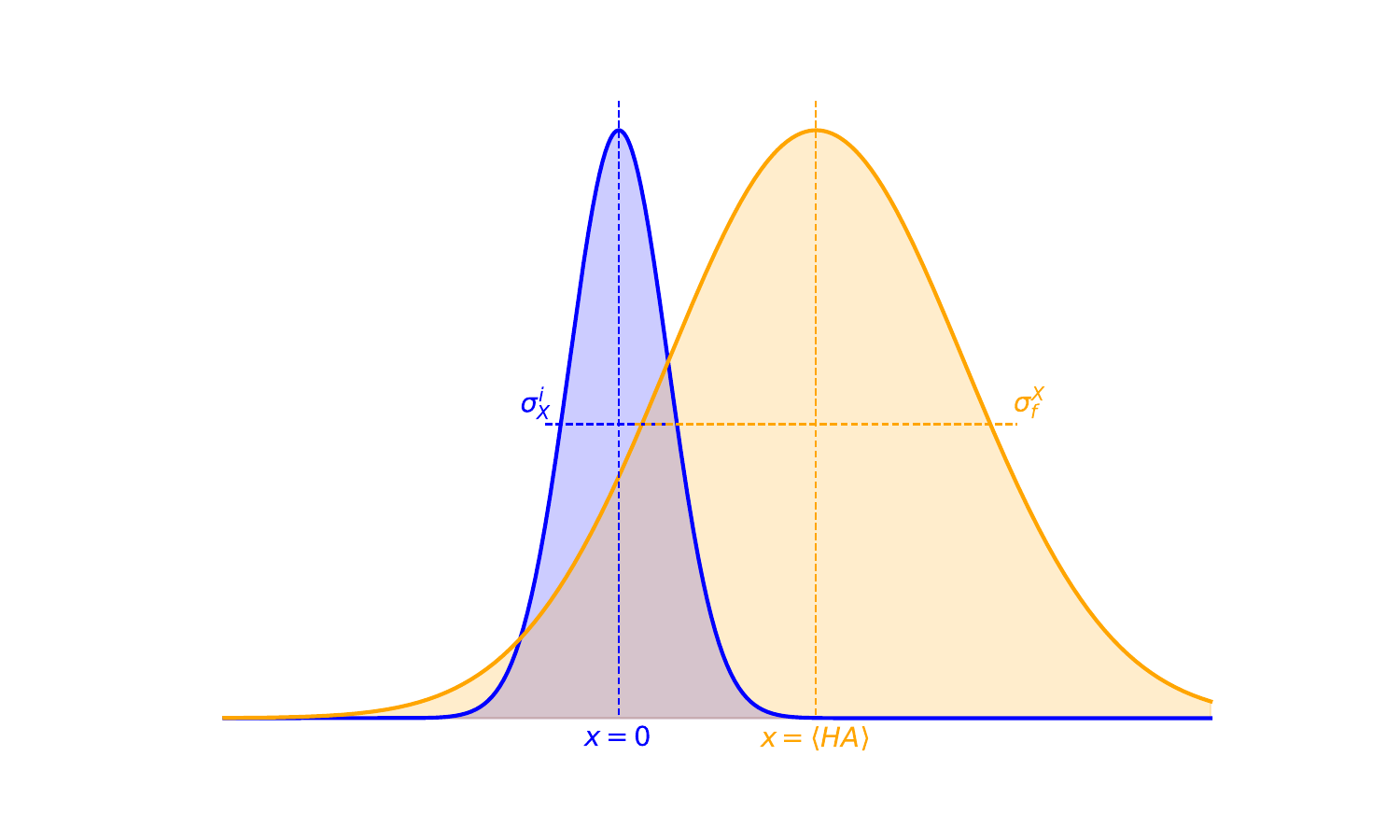}
    \caption{The probability distribution of the pointer location. Blue line denotes the initial probability distribution and the yellow line denotes the final probability distribution. The initial distribution only depends on the resolution of the detector which can be chosen to be arbitrarily small and this distribution is centered at $X=0$. The final distribution depends on the variance of $\hat{H}\hat{A}$, and can be large compared to $\sigma_X$. The probability that the detector detects an excitation approaches one as $\sigma_X$ approaches zero.}
    \label{fig:distribution}
\end{figure}

Now, we move on to consider the probability to achieve a successful detection. To determine the probability of a successful detection, we have to consider the fluctuations in the operator $\hat{X}$ in both the initial and the final state. As described earlier, we assume that the initial state has fluctuation $\sigma_X$ and the distribution is centered at $X=0$. The fluctuation in the final state is given by\footnote{This can be seen as the variance of sum of independent normally distributed random variables that is given by sum of the square of the two variances.}
\begin{equation}
    \begin{split}
        (\sigma_X^f)^2 &= \langle \psi_f| \hat{X}^2 | \psi_f \rangle - \langle \psi_f| \hat{X} | \psi_f \rangle^2\ \\
        &=  \langle \psi_i| (\hat{X}+\hat{H}\hat{A})^2 | \psi_i \rangle - \langle \psi_i| \hat{X} + \hat{H}\hat{A} | \psi_i \rangle^2\,, \\
        &= \sigma_X^2 + \sigma_{HA}^2\,.\label{eq:sigmafinal}
    \end{split}
\end{equation}
Assuming it obeys a Gaussian distribution for the output of measurement of the pointer location, we can obtain the probability of a successful detection as follows. For a successful detection, we demand that the position of the pointer moves by more than $\sigma_X$. Then, the probability of detection is given by the area under the curve of a the final Gaussian distribution of $\hat{X}$, excluding the region $(-\sigma_X,\sigma_X)$. Clearly, when $\sigma_{X}^{f}\gg \sigma_{X}$ i.e. $\sigma_{HA} \gg \sigma_X$, the probability approaches one (see Fig.~\ref{fig:distribution} for a demonstration). 

More interestingly, if one wants the measurements of $HA$ to be of high precision, then one wants $\sigma_{X}^{f}$ to be small. Then, according to Equ.~(\ref{eq:sigmafinal}), $\sigma_{X}$ is also small. This implies that $\sigma_{P}$ is large by the uncertainty principle. Finally, from Equ.~(\ref{eq:sigmadt}), this means that the probability for a big time shift is larger.  As a result, we can conclude from here that \textit{\textbf{the higher the precision of the measurements is the stronger the backreaction is}}.

In the next sub-sections, we compute the lower bound of $\sigma_{\delta t_{AdS}}$ and the value of $\sigma_{HA}$, and show that the lower bound of $\sigma_{\delta t_{AdS}}$ can be arbitrarily small and the probability of a successful detection is close to one. This will complete the proof that, in the present example, it is possible to detect the bulk excitation by our measurement protocol which exploits non-local effects in quantum gravity.

\subsubsection{The Variance of the Time Shift}\label{sec:smearing}
To study this question, let's consider a typical smearing function
\begin{equation}
	f(t,\rho) = {1\over \sqrt{2\pi }\mu} e^{-{t^2\over 2 \mu^2}} \ {\sqrt{2}\over \sqrt{\pi }\lambda} e^{-{\rho^2\over 2 \lambda^2}}\,,\label{eq:smearing}
\end{equation}
with $\mu,\lambda \ll 1$ which gives us a state with an excitation localized near the origin $t=0,\rho=0$.\footnote{Note that the smearing function Equ.~(\ref{eq:smearing}) is normalized as it satisfies $\int_{-\infty}^{\infty}dt\int_{0}^{\frac{\pi}{2}} d\rho f(t,\rho)=1$ due to the smallness of $\lambda$.} Now we can see that the exponent of Equ.~(\ref{eq:psi}) could be expanded as follows.
\begin{equation}
	\int d\rho d\Omega dt \,  f(\rho,t) \phi(\rho,t,\Omega)  = \sum\limits_{n} f_n g_n \left( a_{n , 0} + a_{n,0}^{\dagger} \right)\,, \label{eq:smear1}
\end{equation}
where we have
\begin{equation}
\begin{split}
& f_n = \int\limits_{-\infty}^{\infty} dt \, {1\over \sqrt{2\pi }\mu} e^{-{t^2\over 2 \mu^2}} \, e^{-i \omega_{n,0} t}  =  e^{-{1\over 2} \omega_{n,0}^2 \mu^2}\,,\\
	& g_n = \int\limits_{0}^{\infty} d\rho \, {\sqrt{2}\over \sqrt{\pi }\lambda} e^{-{\rho^2\over 2 \lambda^2}} \, \left( {1\over N_{n,0} } + O(\rho^2) \right) = {1\over N_{n,0}} + O(\lambda^2)\,.
	\end{split}\label{eq:fg}
\end{equation}
Next, we consider the extrapolated boundary operator
\begin{equation}
\begin{split}
\hat{O}(t,\Omega)&=\lim_{\rho_{c}\rightarrow\frac{\pi}{2}}\cos^{-\Delta}\rho\hat{\phi}(t,\rho_{c},\Omega)\,,\\&=\sum_{n,\vec{l}}\frac{_{2}F_{1}(-n,\Delta+l+n,l+\frac{d}{2},1)}{N_{n,\vec{l}}}\Big[a_{n,\vec{l}}e^{-i\omega_{n,\vec{l}}t}Y_{\vec{l}}(\Omega)+a_{n,\vec{l}}^{\dagger}e^{i\omega_{n,\vec{l}}t}Y_{\vec{l}}^{*}(\Omega)\Big]\,,\\&=\sum_{n,\vec{l}}b_{n,\vec{l}}\Big[a_{n,\vec{l}}e^{-i\omega_{n,\vec{l}}t}Y_{\vec{l}}(\Omega)+a_{n,\vec{l}}^{\dagger}e^{i\omega_{n,\vec{l}}t}Y_{\vec{l}}^{*}(\Omega)\Big]\,,
\end{split}
\end{equation}
for which we defined
\begin{equation}
    b_{n,\vec{l}}=(-1)^{n}\frac{\pi}{\sin[\pi(\frac{d}{2}-\Delta)]}\sqrt{\frac{\Gamma(\Delta+n+l)}{\Gamma(n+1)\Gamma(\Delta+n-\frac{d-2}{2})\Gamma(\frac{d}{2}+l+n)\Gamma^{2}(\frac{d}{2}-n-\Delta)\Gamma^{2}(1-\frac{d}{2}+\Delta)}}\,.
\end{equation}
We can compute the fluctuation of this operator in the state $\ket{\Psi}$
\begin{equation}
\begin{split}
    \sigma_{\hat{O}}^{2}&=\bra{\Psi}\hat{O}(t,\Omega)^2\ket{\Psi}=\bra{0}\hat{O}(t,\Omega)^2\ket{0}\,,\\&=\sum_{n,\vec{l}}b_{n,\vec{l}}^2\bra{0}\Big[a_{n,\vec{l}}e^{-i\omega_{n,\vec{l}}t}Y_{\vec{l}}(\Omega)+a_{n,\vec{l}}^{\dagger}e^{i\omega_{n,\vec{l}}t}Y_{\vec{l}}^{*}(\Omega)\Big]^2\ket{0}\,,\\&=\sum_{n,\vec{l}}b_{n,\vec{l}}^{2}Y_{\vec{l}}(\Omega)Y^{*}_{\vec{l}}(\Omega)\,,
    \end{split}
\end{equation}
which is infinite and to see this infinity it is enough to look at the zero angular momentum $\vec{l}=0$ sector
\begin{equation}
\begin{split}
\sigma^{2}_{\hat{O},\vec{l}=0}&=\sum_{n}b_{n,0}^{2}=\sum_{n}\frac{\pi^2}{\sin^2[\pi(\frac{d}{2}-\Delta)]}\frac{\Gamma(\Delta+n)}{\Gamma(n+1)\Gamma(\Delta+n-\frac{d-2}{2})\Gamma(\frac{d}{2}+n)\Gamma^{2}(\frac{d}{2}-n-\Delta)\Gamma^{2}(1-\frac{d}{2}+\Delta)}\,,\\&=\sum_{n}\frac{\Gamma(\Delta+n)\Gamma(\Delta+n-\frac{d-2}{2})}{\Gamma(n+1)\Gamma(\frac{d}{2}+n)\Gamma^{2}(1-\frac{d}{2}+\Delta)}\,,\\&=\frac{1}{\Gamma^{2}(1-\frac{d}{d}+\Delta)}\sum_{n}\frac{\Gamma(\Delta+n)\Gamma(n+1+\Delta-\frac{d}{2})}{\Gamma(\frac{d}{2}+n)\Gamma(n+1)}\,,
\end{split}
\end{equation}
which is divergent due to the fact that $\Delta=\frac{d}{2}+\sqrt{\frac{d^2}{4}+m^2}\geq\frac{d}{2}$ and to go from the first row to second row we used the Euler's reflection formula
\begin{equation}
    \Gamma(1-x)\Gamma(x)=\frac{\pi}{\sin(\pi x)}\,.
\end{equation}
This result is problematic in practice as it tells us that the backreaction caused by measuring the correlator $\bra{\Psi}\hat{H}\hat{O}(t,\Omega)\ket{\Psi}$ is infinitely large. This statement comes from the uncertainty principle which tells us that
\begin{equation}
\begin{split}
\sigma_{\delta t_{AdS}}\sim \frac{\sigma_{\hat{O}}}{|\bra{\Psi}\hat{H}\hat{O}(t,\Omega)\ket{\Psi}|}>\frac{\sigma_{\hat{O},\vec{l}=0}}{|\bra{\Psi}\hat{H}\hat{O}(t,\Omega)\ket{\Psi}|}=\frac{\sigma_{\hat{O},\vec{l}=0}}{|\alpha\sum_{n}f_{n}g_{n}b_{n,0}|}=\infty\,,
\end{split}
\end{equation}
and it is infinite due to the fact that the denominator is finite by the exponential damping of $f_{n}$ in Equ.~(\ref{eq:fg}). Nevertheless, this is just a manifestation of the Bohr-Rosenfeld effect \cite{1933KDVS...12....3B,ROSENFELD1996211,Hubeny:2000eu} which says that the correct observables for a quantum field theory should be smeared local operators. 

Hence we should consider smeared versions of $\hat{O}(t,\Omega)$. For simplicity let's just consider the s-wave smearing and the higher angular-momentum modes can be tamed similarly:
\begin{equation}
	\hat{A} = \lim\limits_{\rho \rightarrow {\pi\over 2}}\cos^{-\Delta}(\rho) \int dt d\Omega F(t) \phi(\rho,t,\Omega) \,,\quad\text{where }
	F(t)  = {1\over \sqrt{2\pi }\sigma} e^{-{t^2\over 2 \sigma^2}}\,.\label{eq:smearingA}
\end{equation}
This smeared boundary operator can be expressed as 
\begin{equation}
	\begin{split}
	\hat{A} &=   \sum\limits_{n} F_n  \left( a_{n , 0} + a_{n,0}^{\dagger} \right)\,, \\
	F_n &= {1\over N_{n,0}} \frac{\pi \Gamma({d\over 2})  \,  e^{-{1\over 2} \omega_{n,0}^2 \sigma^2}}{\sin(\pi({d\over 2} - \Delta)) \Gamma(n+{d\over 2}) \Gamma(\Delta + 1 - {d\over 2}) \Gamma({d\over2} - n - \Delta) }\,.\label{eq:smear2}
	\end{split}
\end{equation}
To detect the, more precisely the s-wave part of the, unitary excitation from the boundary, we measure the following operator.
\begin{equation}
	\begin{split}
	\langle\Psi | \hat{H}\hat{A} |\Psi\rangle &= i\alpha \sum\limits_{n , n'} \omega_{n,0} f_n g_n F_{n'}  \langle 0 |\left( - a_{n , 0} + a_{n,0}^{\dagger} \right) \,  \left( a_{n' , 0} + a_{n',0}^{\dagger} \right) | 0 \rangle\,, \\
	& = - i\alpha \sum\limits_{n } \omega_{n,0} f_n g_n F_{n} \,.
	\end{split}
\end{equation}
The variance of the boundary time evolution caused by this measurement is 
\begin{equation}\label{bulk_time}
	\sigma_{\delta t_{\text{bulk}}}  \sim {\sigma_{\hat{A}} \over |\langle\Psi | \hat{H}\hat{A} |\Psi\rangle  |} \,,
\end{equation}
for which now we have,
\begin{equation}
	\begin{split}
	\sigma_{\hat{A}}^2 &= \langle \Psi | \hat{A}^2 | \Psi \rangle=\langle 0 | \hat{A}^2 |0\rangle  = \sum\limits_{n} F_n^2  = \sum_{n}p(n) e^{-{(2n+\Delta)^2 \sigma^2}}\,,
	\end{split}
\end{equation}
and the summand at large values of $n$ can be estimated as
\begin{equation}
p(n) e^{-{(2n+\Delta)^2 \sigma^2}} \sim  n^{2\Delta - 2d - 2} e^{-{(2n+\Delta)^2 \sigma^2}}\,,
\end{equation}
so we can approximate the summation by an integral
\begin{equation}
\sum_{n} p(n)e^{-(2n+\Delta)^2\sigma^2}\sim \int dx \frac{1}{x}p(x)e^{-(2x+\Delta)^2\sigma^2} \sim {1\over\sigma^{2\Delta-2d - 2}}\,,
\end{equation}
in the limit when $\sigma$ is small. Similarly, we also have
\begin{equation}
	\begin{split}
	\langle\Psi | \hat{H}\hat{A} |\Psi\rangle \sim  \alpha \sum_{n} n^{\Delta - d } e^{ - {1\over 2} {(2n+\Delta)^2 (\sigma^2 + \mu^2)}} \sim {1\over (\sigma^2 + \mu^2)^{\Delta - d \over 2}}\,.
	\end{split}
\end{equation}
Therefore, the variance of the time shit is 
\begin{equation}
    \sigma_{\delta t_{AdS}}= {(\sigma^2 + \mu^2)^{\Delta - d \over 2} \over \sigma^{\Delta-d -1}}=\sigma(1+\frac{\mu^2}{\sigma^2})^{\Delta-d}=\sigma (1+\frac{\mu^2}{\sigma^2})^{\frac{1}{2}\sqrt{d^2+4m^2}-\frac{d}{2}}\,,\label{eq:dp}
\end{equation}
which is small as long as $\frac{\mu}{\sigma}$ is small enough i.e. if the bulk excitation is sufficiently localized in time.

\subsubsection{Measurement Result}
To understand what physical information we can efficiently extract from the measurement, we have to compute both $\bra{\Psi} \hat{H}\hat{A}\ket{\Psi}$ and its variance 
\begin{equation}
\sigma_{HA}=\sqrt{\bra{\Psi}\hat{H}\hat{A}\hat{H}\hat{A}\ket{\Psi}-\bra{\Psi}\hat{H}\hat{A}\ket{\Psi}\bra{\Psi}\hat{H}\hat{A}\ket{\Psi}}\,.
\end{equation}
Using Equ.~(\ref{eq:smear1}), Equ.~(\ref{eq:fg}), Equ.~(\ref{eq:smearingA}) and Equ.~(\ref{eq:smear2}) we have
\begin{equation}
\frac{\sigma^{2}_{HA}}{\bra{\Psi}\hat{H}\hat{A}\ket{\Psi}^2}=1+\frac{\Big(\sum_{n}\omega_{n}^2f_{n}^2g_{n}^2+\alpha^2\sum_{n}\omega_{n}f_{n}^2g_{n}^2\sum_{m}\omega_{m}f_{m}^{2}g_{m}^2\Big)\sum_{l}F_{l}^2+\sum_{n}\omega_{n}f_{n}^{2}g_{n}^2\sum_{m}\omega_{m}F_{m}^{2}}{\sum_{n}\omega_{n}f_{n}g_{n}F_{n}\sum_{m}\omega_{m}f_{m}g_{m}F_{m}}\,.
\end{equation}
This is a rather large number due to the fact that $F_{n}$ is positive and negative alternatively as we vary $n$ but its magnitude is growing with $n$, so there is some cancellations among the contributions to denominator from each $n$ or $m$ but there is no such cancellation for the numerator. As an example for $d=4$ and the scalar mass $m=0.1$ with $\sigma=10^{-14}$ and $\mu=10^{-16}$ we have
\begin{equation}
    \frac{\sigma^{2}_{HA}}{\bra{\Psi}\hat{H}\hat{A}\ket{\Psi}^2}> 2688.85.
\end{equation}
Hence the fluctuation of the $\hat{H}\hat{A}$ measurement is rather large and the precise value of $\bra{\Psi}\hat{H}\hat{A}\ket{\Psi}$ can not be efficiently extracted. This tells us that the final distribution in Fig.\ref{fig:distribution} is very flat. 

As a result, we can only efficiently detect the existence of the excitation in the bulk following our protocol but we cannot efficiently reconstruct its profile as we cannot efficiently determine the value of $\bra{\Psi}\hat{H}\hat{A}\ket{\Psi}$ which is the smeared version of $\bra{\Psi}\hat{H}\hat{O}(t,\Omega)\ket{\Psi}$.\\\\
\noindent\textbf{How is this more than the Gauss Law?}
A critical reader may wonder how is the above protocol different from measuring the energy by utilizing the gravitational Gauss law. This critique is exacerbated by the fact that in this setup, the bulk excitation is dressed to the boundary and can be detected by simply measuring the energy of the system. 

An argument against this critique has been provided in \cite{Laddha:2020kvp}. Consider a situation where there are multiple fields with the same mass.\footnote{To avoid global symmetry, we can consider differences in the mass that are suppressed and can be ignored for the purpose of this perturbative calculation.} The protocol discussed above would enable a boundary observer to detect the type of particle excited, something which is not possible by just using the Gauss law.

\section{Dress the Particle to State-- Situations where Our Protocol Fails}\label{sec:no holography}
In this section, we provide exemplified situations where our previous protocol of particle detection fails to work. These are the examples where the perturbative version of holography proposed in \cite{Chowdhury:2021nxw} fails. We also provide constructions where approximate notions of locality in the bulk exist. To properly establish the results, we will use the Hamiltonian constraints in the canonical quantization of gravity \cite{Arnowitt:1962hi}.

\subsection{Hamiltonian Constraints in AdS}\label{sec:reviewH}
Before we discuss situations where our protocol breaks down, let's establish the precise theoretical tool used in our analysis -- the quantum version of the Hamiltonian constraints in gravity. In the next subsection, we use this tool to construct situations where our particle detecting protocol breaks down. Readers familiar with this tool can jump directly to Sec.~\ref{sec:sdo}.

In quantum gravity, we either study the theory in a fully diffeomorphism-invariant way by quotienting the gauge redundancy from the get-go or we fix the gauge and impose the diffeomorphism constraints by hand afterwards. The first approach is suited to the path integral formalism to the computation of transition amplitudes or partition functions and the second approach fits to the canonical formalism to the construction of the (perturbative) Hilbert space $\mathcal{H}_{\text{pert}}$ and physical operators acting on $\mathcal{H}_{\text{pert}}$. We shall follow the second approach in this paper.

In this subsection, we provide details of the Hamiltonian constraint equations in a background AdS$_{d+1}$ spacetime. We analyze the constraint equations to first order in the gravitational backreaction\footnote{That is we ignore the graviton contribution to the gravitational energy. This approximation is valid in our paper as we only care about bulk excitations created by matter fields. Though our conclusion persists even if graviton is included which requires us to go to the next order in the gravitational backreaction.} which is enough to show that gravitational effects allow us to detect a bulk excitation instantaneously (i.e. only use the information on the same Cauchy slice as the excitation) from the asymptotic boundary using the protocol we developed in this paper.

Let's start with the ADM decomposition of the metric \cite{Arnowitt:1962hi} on a (d+1)-dimensional spacetime
\begin{equation}
ds^{2}=-N^{2} dt^{2}+g_{ij}(dx^{i}+N^{i}dt)(dx^{j}+N^{j}dt)\,,\label{eq:ADM}
\end{equation}
where N is called the \textit{lapse function}, the vector $N^{i}$ is called the \textit{shift vector} and this can be thought of as a general gauge fixing procedure fixing the $d+1$-coordinate reparametrizations. Using the Gauss-Codazzi equation, the matter-coupled Einstein-Hilbert action can be written as
\begin{equation}
S=\frac{1}{16\pi G_{N}}\int dt d^{d}x N\sqrt{g}(R[g]-2\Lambda+K_{ij}K^{ij}-K^{2})+S_{\text{matter}}+S_{\text{bdy}}\,,\label{eq:actionADM}
\end{equation}
where $\Lambda=-\frac{d(d-1)}{2}$ is the cosmological constant, $R[g]$ is the Ricci scalar of the spatial metric $g_{ij}$, $S_{\text{matter}}$ is the action of the minimally coupled matter field which in our case is a free massive scalar field, $S_{\text{bdy}}$ denotes the boundary term of the action which ensures a well-defined variational principle and $K_{ij}$ is the extrinsic curvature of the constant-$t$ slices. Using the metric Equ.~(\ref{eq:ADM}) we can see that the extrinsic curvature is given by
\begin{equation}
K_{ij}=\frac{1}{2N}(-\dot{g}_{ij}+D_{j}N_{i}+D_{i}N_{j})\,,\label{eq:K}
\end{equation}
where $D_{i}$ is the torsion-free and metric-compatible covariant derivative with respect to $g_{ij}$. From Equ.~(\ref{eq:actionADM}) and Equ.~(\ref{eq:K}), we can see that the canonical momenta associated with the lapse function and shift vector are zero
\begin{equation}
\Pi=\frac{1}{\sqrt{g}}\frac{\delta S}{\delta \dot{N}}=0\,,\quad \Pi_{i}=\frac{1}{\sqrt{g}}\frac{\delta S}{\delta \dot{N^{i}}}=0\,.\label{eq:primary}
\end{equation}
Moreover, the Hamiltonian of the system can be written as
\begin{equation}
H_{\text{tot}}=\int d^{d}x\sqrt{g}\Big[N\mathcal{H}+N^{i}\mathcal{H}_{i}\Big]+H_{\text{bdy}}\,,\label{eq:Htot}
\end{equation}
where
\begin{equation}
\begin{split}
  \mathcal{H}&=16\pi G_{N}\Big(\Pi_{ij}\Pi^{ij}-\frac{1}{d-1}(\Pi^{i}_{i})^{2}\Big)-\frac{1}{16\pi G_{N}}(R[g]-2\Lambda)+\mathcal{H}_{\text{matter}}\,,\\
\mathcal{H}_{i}&=-2g_{ij}D_{k}\Pi^{jk}+\mathcal{H}_{i,\text{matter}}\,,\label{eq:constraints}
  \end{split}
\end{equation}
in which $\mathcal{H}_{\text{matter}}$ is the Hamiltonian density of the matter field and $\mathcal{H}_{i,\text{matter}}$ is the momentum density of the matter fields and we have the canonical momentum of $g_{ij}$
\begin{equation}
\Pi^{ij}=\frac{1}{\sqrt{g}}\frac{\delta S}{\delta \dot{g}_{ij}}=-\frac{1}{16\pi G_{N}}\Big(K^{ij}-g^{ij}K\Big)\,.\label{eq:Pigraviton}
\end{equation}
The equations in Equ.~(\ref{eq:primary}) are called \textit{primary constraints} \cite{dirac2001quantum}, which have to be preserved under the time-evolution generated by the Hamiltonian Equ.~(\ref{eq:Htot}). These additional requirements generate the \textit{secondary constraints}
\begin{equation}
\mathcal{H}=0\,,\quad \mathcal{H}_{i}=0\,,
\end{equation}
which constrain physical states and observables after we promote $\mathcal{H}$ and $\mathcal{H}_{i}$ to operators $\hat{\mathcal{H}}$ and $\hat{\mathcal{H}}_{i}$. Physical states in the Hilbert space of the system have to be annihilated by $\hat{\mathcal{H}}$ and $\hat{\mathcal{H}}_{i}$ and gauge invariant observables have to commute with them.

In our paper, we are mostly interested in the constraint $\mathcal{H}=0$ which is called the \textit{Hamiltonian constraint}. As an example to demonstrate how this formalism works, we can analyze it perturbatively around the AdS metric Equ.~(\ref{eq:background}) in which $N=\frac{1}{\cos\rho}$, $N_{i}=0$ and the background metric $g_{ij}^{0}$ is given by
\begin{equation}
g_{ij}^{0}dx^{i}dx^{j}=\frac{d\rho^{2}+\sin^{2}\rho d\Omega^{2}_{d-1}}{\cos^{2}\rho}\,.\label{eq:bc}
\end{equation}
Since the lapse $N$ and shift $N_{i}$ are constrained to be nondynamical, we only have to consider the dynamics from the metric $g_{ij}$. We treat the metric $g_{ij}$ perturbatively and expand it as
\begin{equation}
g_{ij}=g_{ij}^{0}+\sqrt{16\pi G_{N}}h_{ij}\,,
\end{equation}
where $h_{ij}$ is the graviton field which has a normalized kinetic term from expanding the Einstein-Hilbert action to quadratic order in $h_{ij}$.

To the leading order in $G_{N}$ we have 
\begin{equation}
\mathcal{H}^{(0)}=-\frac{1}{16\pi G_{N}}(R[g^{0}]-2\Lambda)=0\,,
\end{equation}
which is constantly true using the geometry Equ.~(\ref{eq:bc}) and the fact that $\Lambda=-\frac{d(d-1)}{2}$. Hence the Hamiltonian constraint is nontrivial starting from the first order, where we have
\begin{equation}
\mathcal{H}^{(1)}=-\frac{1}{\sqrt{16\pi G_{N}}}\Big[(d-1)h+\nabla_{i}\nabla_{j}h^{ij}-\nabla^{2}h\Big]+\mathcal{H}_{\text{matter}}=0\,,\label{eq:core}
\end{equation}
in which $h$ is the trace of $h_{ij}$ and $\nabla_{i}$ is the torsion free and metric compatible covariant derivative with respect to the metric Equ.~(\ref{eq:bc}). From Equ.~(\ref{eq:core}) we have
\begin{equation}
\begin{split}
\sqrt{16\pi G_{N}}H_{\text{matter}}&=\sqrt{16\pi G_{N}}\int d^{d}x
\sqrt{g^{0}}N\mathcal{H}_{\text{matter}}=\int d^{d}x
\sqrt{g^{0}}N\Big[(d-1)h+\nabla_{i}\nabla_{j}h^{ij}-\nabla^{2}h\Big]\,,\\&=\int d^{d}x
\sqrt{g^{0}}\nabla_{i}\Big[\frac{\nabla_{j}h^{ij}-\nabla^{i}h}{\cos\rho}+\frac{\sin\rho}{\cos^{2}\rho}h\delta^{i}_{\rho}-\frac{\sin\rho}{\cos^{2}\rho}h^{i\rho}\Big]\,,\\&=\int d^{d}x\sqrt{g^{0}}\nabla_{i}\Bigg[N\Big[\nabla_{j}h^{ij}-\nabla^{i}h+\tan\rho (h\delta^{i\rho}- h^{i\rho})\Big]\Bigg]\,,\label{eq:HADM}
\end{split}
\end{equation}
which is just a boundary term due to the Stokes theorem. In the usual study of holography, the bulk metric fluctuation or the graviton field satisfies Dirichlet boundary condition near the asymptotic boundary so the last two terms in the above expression are just zero. With this fact incorporated, we can see that the right-hand side is exactly given by the ADM energy as $\sqrt{16 \pi G_{N}}\hat{H}_{\text{ADM}}$ for asymptotic AdS spacetimes \cite{Arnowitt:1962hi,Hawking:1995fd,Giddings:2018umg}. In our main text in Sec.~\ref{sec:no locality} we've called this boundary integral $\sqrt{16\pi G_{N}}\hat{H}$. For the sake of convenience, we will define
\begin{equation}
    \hat{H}_{\partial}=\sqrt{16\pi G_{N}}\hat{H}_{\text{ADM}}\,,
\end{equation}
which is in fact what the boundary observer can measure by measuring metric fluctuations near the boundary. As we have discussed, physical (gauge invariant) operators $\hat{O}(x)$ should commute with the constraint $\hat{\mathcal{H}}$ which to the first order in $G_{N}$ is
\begin{equation}
    [\sqrt{16\pi G_{N}}\hat{H}_{\text{matter}}-\hat{H}_{\partial},\hat{O}(x)]=0\,,\label{eq:commutator}
\end{equation}
which implies that
\begin{equation}
[\hat{H}_{\partial},\hat{O}(x)]=\sqrt{16\pi G_{N}}[\hat{H}_{\text{matter}},\hat{O}(x)]=-i\sqrt{16\pi G_{N}} \frac{\partial}{\partial t}\hat{O}(x)\,.\label{eq:constraintO}
\end{equation}

This is essentially what we did in Sec.~\ref{sec:no locality}. We can see that by measuring the change in the boundary metric we are able to detect a bulk excitation (created by gauge invariant operators) using the leading order effect in $G_{N}$. Moreover, we can see that if we send $G_{N}\rightarrow0$ we are not able to detect anything by measuring boundary metric as it doesn't change. This emphasizes the gravitational nature of our protocol in the main text. In fact, explicit solutions of Equ.~(\ref{eq:constraintO}) can be constructed using gravitational Wilson lines \cite{Donnelly:2018nbv,Giddings:2018umg}.

The momentum constraint $\mathcal{H}_{i}=0$ can be treated similarly. However, it is easy to see from Equ.~(\ref{eq:constraints}) that the momentum constraint $\hat{H}_{i}$ is just the generator of the spatial diffeomorphisms $\epsilon^{i}(x^{j})$. Therefore, this constraint can be easily imposed by making sure that operators are invariant under such diffeomorphisms and we will not analyze it in detail.

\subsection{Dressing to the Features of the Background}\label{sec:sdo}
From Sec.~\ref{sec:reviewH} we see that we can efficiently extract information from the bulk at the boundary, by leveraging  the Hamiltonian constraint Equ.~(\ref{eq:Htot}) in the perturbative regime. This requires us to start with a consistent background configuration and perturbatively expand the Hamiltonian constraint around this background configuration. A consistent background configuration $(g^{0},\phi^{0})$ solves all equations of motion including the matter-coupled Einstein's field equations, which ensures the constraints are satisfied at the zeroth order in the perturbative expansion. The analysis in Sec.~\ref{sec:no locality} is essentially done in the background configuration that $g^{0}$ is the metric for an empty AdS$_{d+1}$ spacetime with no background matter distribution, i.e. $\phi^{0}=0$.

Now let's consider the case of a background with a nontrivial matter distribution $\phi^{0}(x)$ and the background geometry is still in the form Equ.~(\ref{eq:ADM}). In this background, the free massive scalar matter field has the action
\begin{equation}
\begin{split}
S&=-\frac{1}{2}\int d^{d+1}x\sqrt{-G}\Big[G^{\mu\nu}\partial_{\mu}\phi(x)\partial^{\nu}\phi(x)+m^{2}\phi^{2}(x)\Big]\,,\\&=\frac{1}{2}\int d^{d+1}x\sqrt{g}N\Big[\frac{1}{N^2}\Big(\dot{\phi}(x)-N^{i}\partial_{i}\phi(x)\Big)^{2}-g^{ij}\partial_{i}\phi(x)\partial_{j}\phi(x)-m^{2}\phi^{2}(x)\Big]\,,
\end{split}
\end{equation}
where we use $G_{\mu\nu}$ to denote the spacetime metric with the lapse and shift included and $g_{ij}$ to denote the spatial metric in the ADM decomposition Equ.~(\ref{eq:ADM}). The canonical momentum of the scalar field is given by
\begin{equation}
\pi(x)=\frac{1}{\sqrt{g}}\frac{\delta S}{\delta\dot{\phi}(x)}=\frac{1}{N}(\dot{\phi}(x)-N^{i}\partial_{i}\phi(x))\,,
\end{equation}
which satisfies the equal-time canonical commutation relation with the field operator $\hat{\phi}(x)$
\begin{equation}
    [\hat{\pi}(t,\vec{x}),\hat{\phi}(t,\vec{y})]=-i\frac{1}{\sqrt{g}}\delta^{d}(\vec{x}-\vec{y})\,.
\end{equation}
Now we can read out the matter Hamiltonian as
\begin{equation}
\begin{split}
    H_{\text{matter}}&=\int d^{d}x\sqrt{-G} \Big[\pi(x) \dot{\phi}(x)-\mathcal{L}\Big]\,,\\&=\int d^{d}x\sqrt{g}\Big[\frac{N}{2}\Big(\pi^{2}(x)+g^{ij}\partial_{i}\phi(x)\partial_{j}\phi(x)+m^{2}\phi^{2}(x)\Big)+N^{i}\pi(x)\partial_{i}\phi(x)\Big]\,,
\end{split}
\end{equation}
from which we can see that the matter contribution to the secondary constraints Equ.~(\ref{eq:constraints}) are
\begin{equation}
\begin{split}
\mathcal{H}_{\text{matter}}&=\frac{1}{2}\Big(\pi^{2}+g^{ij}\partial_{i}\phi\partial_{j}\phi+m^{2}\phi^{2}\Big)\,,\\\mathcal{H}_{i,\text{matter}}&=\pi\partial_{i}\phi\,.
\end{split}
\end{equation}
Now we want to study the Hamiltonian constraint by linearizing it around a consistent background $(g^{0},\phi^{0})$. We will consider the expansion
\begin{equation}
g_{ij}=g_{ij}^{0}+\sqrt{16\pi G_{N}}h_{ij}\,,\quad \phi(x)=\phi^{0}(x)+\Phi(x)\,,
\end{equation}
and we will work exactly in $\Phi(x)$ but only to first nontrivial order in $G_{N}$. The conjugate momentum for $\Phi(x)$ can be easily worked out as before
\begin{equation}
\pi_{\Phi}=\frac{1}{N}(\dot{\phi}_{0}-N^{i}\partial_{i}\phi^{0}+\dot{\Phi}-N^{i}\partial_{i}\Phi)\,,
\end{equation}
which satisfies the equal-time canonical commutation relation with $\hat{\Phi}(x)$ when promoted to an operator\footnote{Similar results are obtained in \cite{Giddings:2022hba,Giddings:2022sss}.}
\begin{equation}
\begin{split}
[\hat{\pi}_{\Phi}(t,\vec{x}),\hat{\Phi}(t,\vec{y})]&=[\frac{1}{N}(\dot{\phi}_{0}-N^{i}\partial_{i}\phi^{0}+\dot{\hat{\Phi}}-N^{i}\partial_{i}\hat{\Phi})(t,\vec{x}),\hat{\Phi}(t,\vec{y})]\,,\\&=[\frac{1}{N}(\dot{\hat{\Phi}}-N^{i}\partial_{i}\hat{\Phi})(t,\vec{x}),\hat{\Phi}(t,\vec{y})]\,,\\&=-i\frac{1}{\sqrt{g}}\delta^{d}(\vec{x}-\vec{y})\,.
\end{split}
\end{equation}
The result of the linearization is the same as before in Equ.~(\ref{eq:core}) with an additional term from linearizing $16\pi G_{N}\Big(\Pi_{ij}\Pi^{ij}-\frac{1}{d-1}(\Pi^{i}_{i})^{2}\Big)$ in this new background
\begin{equation}
\begin{split}
\mathcal{H}^{(1)}&=-\frac{1}{\sqrt{16\pi G_{N}}}\Big[(d-1)h+\nabla_{i}\nabla_{j}h^{ij}-\nabla^{2}h\Big]+\mathcal{H}_{\text{matter}}(\Phi)\\&+2\sqrt{16\pi G_{N}}\Big(\Pi_{ij}^{0}\pi^{ij}-\frac{1}{d-1}\Pi^{i0}_{i}\pi^{j}_{j}\Big)\,,\label{eq:core2}
\end{split}
\end{equation}
where we denote the $\Phi$ relevant part of the $\mathcal{H}_{\text{matter}}$ by $\mathcal{H}_{\text{matter}}(\Phi)$ with $\Pi_{ij}^{0}$ the background value of Equ.~(\ref{eq:Pigraviton}). Furthermore, the equation of motion and canonical quantization condition tells us that
\begin{equation}
[\hat{\mathcal{H}}_{\text{matter}}(\Phi)(t,\vec{x}),\hat{\Phi}(t,\vec{y})]=-i\frac{1}{\sqrt{g^{0}}N}\Big(\dot{\phi^0}(t,\vec{x})-N^{i}\partial_{i}\phi^0(t,\vec{y})+\dot{\hat{\Phi}}(t,\vec{x})-N^{i}\partial_{i}\hat{\Phi}(t,\vec{y})\Big)\delta^{d}(\vec{x}-\vec{y})\,,
\end{equation}
and
\begin{equation}
    [\pi^{ij}(t,\vec{x}),h_{lm}(t,\vec{y})]=-i\delta^{(i}_{(l}\delta^{j)}_{m)}\frac{1}{\sqrt{g^{0}}N}\delta^{d}(\vec{x}-\vec{y})\,.\label{eq:gravitoncanonical}
\end{equation}

Now the integrated Hamiltonian constraint Equ.~(\ref{eq:core2}) is
\begin{equation}
\begin{split}
    \int d^{d}x\sqrt{g^{0}}N \mathcal{H}^{(1)}&=\int d^{d}x\sqrt{g^{0}}N\Bigg[-\frac{1}{\sqrt{16\pi G_{N}}}\Big[(d-1)h+\nabla_{i}\nabla_{j}h^{ij}-\nabla^{2}h\Big]+\mathcal{H}_{\text{matter}}(\Phi)\Bigg]\\&\quad\quad\quad\quad+2\sqrt{16\pi G_{N}}\int d^{d}x\sqrt{g^{0}}N\Big(\Pi_{ij}^{0}\pi^{ij}-\frac{1}{d-1}\Pi^{i0}_{i}\pi^{j}_{j}\Big)\,,\\&=\int d^{d}x\sqrt{g^{0}}\Bigg[-\frac{1}{\sqrt{16\pi G_{N}}}\Big[(d-1)Nh-\nabla_{i}N\nabla_{j}h^{ij}+\nabla_{i}N\nabla^{i}h\\&\quad\quad\quad\quad\quad+\nabla_{i}(N\nabla_{j}h^{ij})-\nabla_{i}(N\nabla^{i}h)\Big]+\mathcal{H}_{\text{matter}}(\Phi)\Bigg]\\&\quad\quad\quad\quad+2\sqrt{16\pi G_{N}}\int d^{d}x\sqrt{g^{0}}N\Big(\Pi_{ij}^{0}\pi^{ij}-\frac{1}{d-1}\Pi^{i0}_{i}\pi^{j}_{j}\Big)\,,\\&=-\frac{1}{\sqrt{16\pi G_{N}}}H_{\partial}+\int d^{d}x\sqrt{g^{0}}N\mathcal{H}_{\text{matter}}(\Phi)\\&\quad\quad-\frac{1}{\sqrt{16\pi G_{N}}}\int d^{d}x\sqrt{g^{0}}\Big[(d-1)Nh-\nabla_{i}N\nabla_{j}h^{ij}+\nabla_{i}N\nabla^{i}h\Big]\\&\quad\quad\quad\quad+2\sqrt{16\pi G_{N}}\int d^{d}x\sqrt{g^{0}}N\Big(\Pi_{ij}^{0}\pi^{ij}-\frac{1}{d-1}\Pi^{i0}_{i}\pi^{j}_{j}\Big)\,,\\&\equiv-\frac{1}{\sqrt{16\pi G_{N}}}H_{\partial}+\int d^{d}x\sqrt{g^{0}}N\mathcal{H}_{\text{matter}}(\Phi)+H_{\text{bulk}}(h)\,.\label{eq:ham}
\end{split}
\end{equation}
This constraint will constrain operators $\hat{O}(x)$ defined in the background $(g^0,\phi^0)$ as 
\begin{equation}
\left[-\frac{1}{\sqrt{16\pi G_{N}}}\hat{H}_{\partial}+\int d^{d}x\sqrt{g^{0}}N\hat{\mathcal{H}}_{\text{matter}}(\Phi)+\hat{H}_{\text{bulk}}(h),\hat{O}(x)\right]=0\,.\label{eq:co}
\end{equation}
In this background $(g^0,\phi^0)$ we have a chance to construct operators $\hat{O}(x)$ satisfying the equation Equ.~(\ref{eq:co}) and also commuting with the ADM Hamiltonian $\hat{H}_{\text{ADM}}=\frac{1}{\sqrt{16\pi G_{N}}}\hat{H}_{\partial}$. The idea is to dress the operator to the background $(g^{0},\phi^{0})$. As a concrete example, we can consider the following operator that is defined such that its position is defined with respect to some background expectation value, 
\begin{equation}
\hat{O}(a)=\int d^{d+1}y\sqrt{g^0+\sqrt{16\pi G_{N}}\hat{h}}N(\hat{\Phi}(y)+\phi^{0}(y))\delta(\phi^{0}(y)+\hat{\Phi}(y)-a)\,,\label{eq:backgrounddressed}
\end{equation}
where the operator $\hat{\Phi}(y)$ inside the delta-function should be understood using Taylor expansion of the delta-function around $\phi^{0}(y)-a$. Using Equ.~(\ref{eq:core2}) and Equ.~(\ref{eq:gravitoncanonical}) it is easy to see that this operator satisfies Equ.~(\ref{eq:co}) at leading order in $G_{N}$, for which the $\sqrt{16\pi G_{N}}\hat{h}$ term is important as the $H_{\text{bulk}}(h)$ has a prefactor $\frac{1}{\sqrt{16\pi G_{N}}}$. Meanwhile, this operator commutes with the boundary ADM Hamiltonian.

In this case, the bulk excitation created by the unitary operator
\begin{equation}
\hat{U}=e^{i\alpha \int d^{d+1}x\sqrt{g^{0}}N f(x)\hat{O}(x)}\,,
\end{equation}
couldn't be detected by a near-boundary observer using the protocol in Sec.~\ref{sec:no locality}. We notice that for such an operator to be well-defined, i.e. to have a finite operator norm, the background configuration $\phi^{0}(x)$ cannot be zero. On the other hand, the case with a zero background configuration is exactly the situation where our protocol in Sec.~\ref{sec:no locality} works. Moreover, we should note that the dressing for this operator makes use of the $\delta$-function which might have multiple spacelike separated supports so in general the resulting operator may not be a truly localized observable. This issue can be remedied by the construction in \cite{Geng:2024dbl}. The construction in \cite{Geng:2024dbl} is motivated by the following observation that when the background matter distribution has enough features we don't need the graviton field as in Equ.~(\ref{eq:backgrounddressed}) for the dressing at all. A preliminary example of such a construction in given in \cite{Giddings:2005id}, where one has $(d+1)$ fields $\hat{Z}^{\mu}(x)$ that have vacuum expectation values $x^{\mu}$ which 
can be used as the coordinates. In this case, we can locally dress an operator $\hat{\phi}(x)$ as
\begin{equation}
    \hat{O}(\xi)=\int d^{d+1}y \hat{\phi}(y)\delta^{d+1}(\hat{Z}^{\mu}(y)-\xi^{\mu})\abs{\det\frac{\partial \hat{Z}(y)}{\partial y}}\,.\label{eq:GHM}
\end{equation}
 In fact, a more general construction can be performed if one realizes that Equ.~(\ref{eq:GHM}) equals to $\hat{\phi}(\xi-\delta\hat{Z}(\xi))$, where $\delta \hat{Z}^{\mu}(x) $ is the fluctuation of $\hat{Z}^{\mu}(x)$ around its vacuum expectation value $x^{\mu}$. Such constructions are streamlined in \cite{Geng:2024dbl} where it is manifest how sharply the operator is localized.\footnote{See \cite{Seo:2025tsw} for some applications of this general construction and \cite{Francois:2024rdm} for some relevant discussions.}

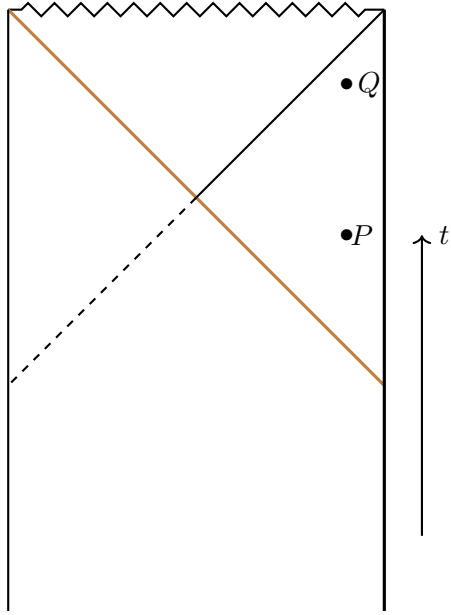
\begin{figure}[h]
    \centering
    \begin{tikzpicture}
       \draw[-, thick] 
       decorate[decoration={zigzag,pre=lineto,pre length=5pt,post=lineto,post length=5pt}] {(-2.5,0) to (2.5,0)};
       \draw[brown,very thick] (-2.5,0) to (2.5,-5);
       \draw[-,very thick,black] (2.5,0) to (2.5,-8);
       \draw[-,thick,black] (2.5,0) to (0,-2.5);
       \draw[dashed,thick,black] (0,-2.5) to (-2.5,-5);
       \draw[thick,black] (-2.5,0) to (-2.5,-8);
       \draw[->,thick,black] (3,-7) to (3,-3);
       \node at (3.3,-3) {\textcolor{black}{$t$}};
       \node at (2,-3) {\textcolor{black}{$\bullet$}};
       \node at (2.2,-3) {\textcolor{black}{$P$}};
       \node at (2,-1) {\textcolor{black}{$\bullet$}};
       \node at (2.3,-1) {\textcolor{black}{$Q$}};
    \end{tikzpicture}
    \caption{The Penrose diagram of a large black hole in AdS formed from collapsing a matter shell. The shell's trajectory is in brown. We take two bulk points $P$ and $Q$ denoting bulk points in early time and late time respectively. The asymptotic time coordinate is denoted as $t$.}
    \label{pic:penroseAdSBH}
\end{figure}

 Before we wrap up the discussion in this subsection, let's make a comment on the relevance of the construction we did to black holes. Let's consider a large black hole in AdS$_{d+1}$ formed from collapsing a matter shell (see Fig.~\ref{pic:penroseAdSBH}). One might wonder if we could use the construction we provided in this subsection to construct operators in the exterior of such a black hole spacetime that perturbatively commutes with the boundary ADM Hamiltonian. In this scenario, one thinks of the shell as the feature of the background. Nevertheless, our construction also requires the background feature has a known description as the vacuum expectation value of some operators and the fluctuating part of such operators is essential for our construction. This fluctuating part is essential as it can be thought of as the collective coordinate of the background feature which can be used to compensate the diffeomorphism transform of the local operators that we are trying to dress \cite{Geng:2024dbl}. Thus, our construction can be done only if the microstructure of the shell is specified.\footnote{We note that this is different from the treatment of the shell as a classical stress-energy tensor source in general relativity. The construction cannot be done in that context as the shell is featureless.} For our purpose, let's denote the operator that sources the shell as $\hat{S}(x)$ which has a nontrivial background configuration $S(s)$ describing the shell where $s$ denotes the affine worldline coordinate of the collapsing shell. The configuration $S(s)$ is spherically symmetric. Thus, a clock operator can be constructed from the shell operator, and the rod operator can be constructed using the geodesic distance from the clock. Given a bulk scalar field $\hat{\phi}(x)$, one can perform the construction as
 \begin{equation}
     \hat{\phi}^{\text{bulk}}(s,\vec{l})=\int d^{d+1}{x}\sqrt{g(x))}N(x)\int d^{d+1}y\sqrt{g(y))}N(y)\hat{\phi}(t_{x},\vec{y})\delta(\hat{S}(x)-S(s))\delta_{g}(\vec{y}-\vec{x}-\vec{l})\,,\label{eq:geodress}
 \end{equation}
where for simplicity we have denoted $g=\det\big(g^{0}+\sqrt{16\pi G_{N}}\hat{h}\big)$ and $\delta_{g}(\vec{y}-\vec{x}-\vec{l})$ denotes that the geodesic distance between each component of $\vec{y}$ and the corresponding component of $\vec{x}$ under the metric $g_{\mu\nu}=g^{0}_{\mu\nu}+\sqrt{16\pi G_{N}}\hat{h}_{\mu\nu}$ equals the corresponding component of $\vec{l}$. Intuitively, the above construction works in the following way that we first use a point on the shell to specify the time and then construct a spatial coordinate with respect to this point. We note that such a construction works nicely at the early times (for example for a local operator at the point $P$ in Fig.~\ref{pic:penroseAdSBH}). Nevertheless, at late times ( $0\ll t< e^{e^{S_{BH}}}$) the shell is deep in the interior of the black hole, and due to the linear growing of the black hole interior in time the distance between an exterior point (for example $Q$ in Fig.~\ref{pic:penroseAdSBH}) and the reference point on the shell is very large. Thus, at late times the above construction would not give an operator sharply supported at a point in the black hole exterior. 

This can be seen more explicitly by noticing that the spacelike geodesic connecting the bulk operator to the feature deep in the black hole interior that it must be dressed to grows at late times, as seen from the Schwarzschild metric. In such a situation, similar to inflating cosmologies, one expects that local metric fluctuations become approximately uncorrelated. This suggests a classical statistical estimate of $\sqrt{\ell \ell_P}$ for the variation of the geodesic distance, $\ell$, between the feature and the target point. We note that this is consistent with  \cite{Verlinde:2019ade,Verlinde:2019xfb} in a different context. So at late times, for example when $t\sim \frac{\beta^2}{l_{P}}$, where we have $l\sim t$ the fluctuation of the location of the geodesic-dressed operator Equ.~(\ref{eq:geodress}) would be much bigger than the size of the black hole. In fact, we expect that the above construction wouldn't give a sensible local operator on the black exterior at a much earlier time $t\sim \beta$. This is because at this  time scale the black hole equilibrates and the exterior observer wouldn't  realize that the black hole is formed from the shell collapsing \cite{Geng:2021hlu}. Thus, our particle detection protocol in Sec.~\ref{sec:no locality} should work in this late-time regime for exterior excitations.\footnote{In fact, we expect our protocol to work until $t\sim e^{S_{BH}}$ where the time stops being a well-defined semiclassical observable as its rank cannot exceeds the dimension of the black hole microstate Hilbert space. Moreover, at the Poincar\'{e} recurrence time $t\sim e^{e^{S_{BH}}}$ the shell remerges from the boundary and starts collapsing again, which transforms the black hole to a white hole. We leave the detailed analysis of the physics associated to these time scales to future work.}
 
\subsection{Approximate Bulk Local Operators-- Hiding a Needle in Haystack}\label{sec:approxlocality}
The analysis in Sec.~\ref{sec:sdo} suggests that a simple bulk operator localized to a non-trivial background configuration with strong enough features makes the operator undetectable by the boundary observer using the protocol we described. Nevertheless, one might want to ask if there is any scenario in between the situations we considered in Sec.~\ref{sec:no locality} and Sec.~\ref{sec:sdo} such that we can have some approximate local operators in the bulk that are hard to detect by a near-boundary observer in a spacetime background which lacks classical strong features. In this subsection we will provide two constructions of such scenario.

Firstly, let's consider the following situation that we have $M$ free massive scalar fields in an empty AdS$_{d+1}$ background (with different masses such that there is no global symmetry). Let's call these scalar fields $\hat{\phi}_{i}$ where $i=1,2,\cdots M$ and denote the corresponding extrapolated operators according to Equ.~(\ref{eq:O}) for each of them as $\hat{\mathcal{O}}_{i}$. If we consider all of these fields to be in the ground state then an approximate notion of locality exists for a coarse-grained boundary algebra which contains the ADM Hamiltonian $\hat{H}_{\text{ADM}}$ and the averaged operator
\begin{equation}
    \hat{O}_{\text{ave}}=\frac{\sum_{i=1,2,\cdots M}\hat{\mathcal{O}}_{i}}{\sqrt{M}}\,.
\end{equation}
We note that the averaged operator $\hat{O}_{\text{ave}}$ has a unit operator norm if the operators $\hat{\mathcal{O}}_{i}$ have unit operator norms. If we take $M$ to be large then our ability to detect a bulk excitation created by the unitary operators
\begin{equation}
    U_{i}=e^{i \alpha \int d\rho d\Omega dt \,  f(\rho,t) \hat{\phi}_{i}(\rho,t,\Omega) }\,, 
\end{equation}
by measuring $\hat{H}\hat{O}_{\text{ave}}$ following the protocol in Sec.~\ref{sec:no locality} is suppressed by a factor $\frac{1}{\sqrt{M}}$. This provides an approximate notion of locality in that it can be difficult for an boundary observer to detect a bulk excitation with such a coarse-grained operator algebra.

Another simple situation in which there is power law suppression of the fidelity of extracting bulk information occurs when there is uncertainty in the energy of the background state. Consider a weakly gravitating bulk quantum field theory in a superposition of energy eigenstates $\ket{\Psi}$ such that the averaged energy $\bra{\Psi}\hat{H}_{\text{matter}}\ket{\Psi}$ is small in Planck unit
\begin{equation}
    \bra{\Psi}\hat{H}_{\text{matter}}\ket{\Psi}\ll G_{N}^{-\frac{1}{d-1}}\,,
\end{equation}
so we can approximate the background geometry to be global AdS$_{d+1}$. Before we proceed we should emphasize that as in Sec.~\ref{sec:sdo2} we are ignoring backreaction as opposed to our discussion in Sec.~\ref{sec:sdo}, however this configuration is close to an on-shell configuration and still approximately satisfies the Hamiltonian constraint. We can choose the state $\ket{\Psi}$ such that we have
\begin{equation}
    \bra{\Psi} \hat{H}_{\text{matter}}^{2}\ket{\Psi}-\bra{\Psi}\hat{H}_{\text{matter}}\ket{\Psi}^2=\Sigma\ll G_{N}^{-\frac{2}{d-1}}\,,
\end{equation}
where we have set the AdS length scale to one. The Hamiltonian constraint forces $\hat{H}_{\text{matter}}=\hat{H}_{\text{ADM}}$. Hence, the unit normalized Hamiltonian operator we should use is 
\begin{equation}
\hat{H}'=\frac{\hat{H}_{\text{matter}}-\bra{\Psi}\hat{H}_{\text{matter}
}\ket{\Psi}}{\sqrt{\Sigma}}\,.\label{eq:normalizedH}
\end{equation}
Moreover, the state $\ket{\Psi}$ is constructed such that in this state the operator $\hat{O}_{1}$ has a unit norm 
\begin{equation}
\bra{\Psi}\hat{O}_{1}\hat{O}_{1}\ket{\Psi}=1.
\end{equation}
In other words, the state $\ket{\Psi}$ is properly chosen such that the matter sector field $\phi_{1}$ is not excited but others are excited. For example, such a state can be a Gaussian distribution of the energy eigenstates of the low-energy matter fields other than $\phi_1$.

Now we can try to detect an excitation created by a bulk operator 
\begin{equation}
    U=e^{i \alpha \int d\rho d\Omega dt \,  f(\rho,t) \hat{\phi}_{1}(\rho,t,\Omega) }\,, 
\end{equation}
following our protocol in Sec.~\ref{sec:no locality} by measuring $\hat{H}'\hat{O}_{1}$ in the excited state $U\ket{\Psi}$
\begin{equation}
\bra{\Psi}U^{\dagger}\hat{H}'\hat{O}_{1} U\ket{\Psi}
\end{equation}
to first order in $\alpha$ on the boundary that is spacelikely separated from the bulk excitation. We get
\begin{equation}
\begin{split}
    &-i\alpha\int d\rho d\Omega dt \,  f(\rho,t) \Big[\bra{\Psi}\hat{\phi}_{1}(\rho,t,\Omega)\hat{H}'\hat{O}_{1}\ket{\Psi}-\bra{\Psi}\hat{H}'\hat{O}_{1}\hat{\phi}_{1}(\rho,t,\Omega)\ket{\Psi}\Big]\,,\\=&-i\alpha\int d\rho d\Omega dt \,  f(\rho,t)\Big[i\frac{\bra{\Psi}\partial_{t}\hat{\phi}_{1}(\rho,t,\Omega)\hat{O}_{1}\ket{\Psi}}{\sqrt{\Sigma}}+\bra{\Psi}\hat{H}'\hat{O}_{1}\hat{\phi}_{1}(\rho,t,\Omega)\ket{\Psi}-\bra{\Psi}\hat{H}'\hat{O}_{1}\hat{\phi}_{1}(\rho,t,\Omega)\ket{\Psi}\Big]\,,\\=&\alpha\int d\rho d\Omega dt \,  f(\rho,t)\frac{\bra{\Psi}\partial_{t}\hat{\phi}_{1}(\rho,t,\Omega)\hat{O}_{1}\ket{\Psi}}{\sqrt{\Sigma}}\,,
    \end{split}
\end{equation}
where in the last step we used the fact that the boundary operator $\hat{O}_{1}$ and the bulk excitation are spacelike separated. Thus, we can see that the result is suppressed by the factor $\frac{1}{\sqrt{\Sigma}}$. Therefore, if we take the following double scaling limit
\begin{equation}
    \Sigma\gg1\,,\quad \Sigma G^{\frac{2}{d-1}}\ll1\,, 
\end{equation}
then we see that the bulk excitation created by the operator $U$ is approximately invisible to the boundary. However, one might wonder what if we just increase the precision of the energy measurement and if one could efficiently operate our protocol in Sec.~\ref{sec:no locality} with this increased precision to detect the bulk excitation created by the operator $U$. The answer is that it would take a longer time for one to operate this protocol as the timescale in Equ.~(\ref{eq:dp}) will be multiplied by how much the precision is increased, i.e. by a factor $\sqrt{\Sigma}$. Thus, our protocol in Sec.~\ref{sec:no locality} becomes less efficient.

\subsection{Approximate Bulk Local Operators-- Emergent Clocks}\label{sec:sdo2}
The discussion in Sec.~\ref{sec:sdo} suggests that if the background has strong enough features then one might be able to dress local bulk operators to these features. More intuitively, one can think of these features as defining a natural clock that a bulk observer can use to approximately describe local physics, for example an actively evolving supernova can be used as an effective clock in the universe.\footnote{Similarly, a distribution of matter with a strong variance of brightness in different positions can be used as a coordinate system. We mainly consider the clocks in this section as we are dealing with the Hamiltonian constraint. The consideration easily generalizes to momentum constraints if one has the above kind of distributions of matter.} Our goal in this subsection is to formulate this intuition in a precise way and hence put the simplified models in \cite{Chandrasekaran:2022cip} to a more concrete footing. As we will see, a relevant example where our construction applies is a typical black hole microstate. 

Since we are working in AdS space whose global patch has compact spatial directions, the exact CFT spectrum  is discrete. Here the intuition is that the inverse of the (typical) level spacing of the energy spectrum appearing in the background states in question is a time scale which can be used as a clock in states with strong enough features. 

Consider a collection of states
\begin{align}
    \ket{f,a}=\sum_{k,i}f_i(E_{k})a_{k}^i\ket{E_{k}}&\approx\sum_{\alpha,i}f_i(E_{\alpha})\sum_{n\in\mathcal{B_\alpha}}a_{n}^i\ket{E_{n}}\,,\label{eq:statewithfeature}
\end{align}
where each $\mathcal{B}_{\alpha}$ denotes a microcanonical band for energy eigenstates. Here $f_i(E)$ are smoothly varying functions of energy, that label several families of configurations, and the $a^i_n$ are the erratic microstate data of those configurations expanded in terms of exact energy eigenstates. We imagine that the semi-classical experimenter can manipulate the $f_i$ not the $a^i_n$.

We then define the following Hermitian operator that functions as a physical clock variable 
\begin{equation}
    \hat{D}=-\frac{i}{2\pi}\sum_{n, m}\frac{\sigma R_{mn} (E_{n}-E_{m})}{[(E_{n}-E_{m})^{2}+\sigma^{2}]^2}\Big[\ket{E_{n}}\bra{E_{m}}-\ket{E_{m}}\bra{E_{n}}\Big]\,,\label{eq:clockoperator}
\end{equation}
where $R_{mn}$ is a random matrix whose statistics will be specified below in a way associated to the given collection of states, and $\sigma$ is a tolerance parameter on the precision of the clock that must be taken to be larger than the level spacing.

Now we have:
\begin{align}
    &\bra{m}[\hat{H}_{\text{matter}},\hat{D}]\ket{n} = -\frac{i}{\pi}\frac{\sigma R_{mn} (E_{n}-E_{m})^2}{[(E_{n}-E_{m})^{2}+\sigma^{2}]^2}\,,\\
\end{align}
Thus, we have the commutator $[H,D]$ evaluated between any two states $\ket{f,a}$ and $\ket{g,a}$: 
\begin{align}
    \bra{g,a}[\hat{H}_{\text{matter}},\hat{D}]\ket{f,a}=&\sum_{\alpha,\beta,k,l} f_{k}(E_\alpha)g_{l}(E_\beta)\sum_{m\in B_{\alpha},n \in B_{\beta}}a^{*l}_{m}a^{k}_n\bra{m}[\hat{H}_{\text{matter}},\hat{D}]\ket{n} \\=&\sum_{\alpha,\beta,k,l} f_{k}(E_\alpha)g_{l}(E_\beta)\sum_{m\in B_{\alpha},n \in B_{\beta}}\Bigg[-\frac{i}{\pi}\frac{\sigma a_m^{l*}a^{k}_nR_{mn} (E_{n}-E_{m})^2}{[(E_{n}-E_{m})^{2}+\sigma^{2}]^2}\Bigg]\,.
\end{align}
We take the average over the random flavor indices. The pseudo-random vectors $a_{n}^{i}$ and the random matrix $R_{mn}$ are taken to obey
\begin{equation}
    \overline{a^{l*}_{n}a_{n'}^{k}}=\delta_{nn'}\delta^{ik}\,,\quad\overline{a_{m}^{l*}R_{m'n'}a^{k}_{n}}=C(E_n, E_m)\delta_{nn'}\delta_{mm'}\delta^{lk}\,,\label{eq:ensemble}
\end{equation}
where the smooth function $C(E,E')$ will be specified below. Thus, we have
\begin{align}
    \overline{\bra{g,a}[\hat{H}_{\text{matter}},\hat{D}]\ket{f,a}}=&\sum_{\alpha,\beta,k} f_{k}(E_\alpha)g_{k}(E_\beta)\sum_{m\in B_{\alpha},n \in B_{\beta}}\Bigg[-\frac{i}{\pi}\frac{\sigma  (E_{n}-E_{m})^2}{[(E_{n}-E_{m})^{2}+\sigma^{2}]^2}\Bigg]\,.
\end{align}
Now let's consider the density of states within each microcanonical band to be $\rho(E)$ with band width $\delta E$. We have
\begin{align}
    \overline{\bra{g,a}[\hat{H}_{\text{matter}},\hat{D}]\ket{f,a}}=&\sum_{\alpha,\beta,k} f_{k}(E_\alpha)g_{k}(E_\beta)\rho(E_{\alpha})\rho(E_{\beta})\delta E_{\alpha}\delta E_{\beta}C_{\alpha\beta}\Bigg[-\frac{i}{\pi}\frac{\sigma  (E_{\alpha}-E_{\beta})^2}{[(E_{\alpha}-E_{\beta})^{2}+\sigma^{2}]^2}\Bigg]\\=&\sum_{k}\int dE dE'\rho(E) f_{k}(E)\rho(E')g_{k}(E')C(E,E')\Bigg[-\frac{i}{\pi}\frac{\sigma  (E-E')^2}{[(E-E')^{2}+\sigma^{2}]^2}\Bigg]\,.
\end{align}
We shall take $\sigma$ to be small enough compared to the size of the bands, so that if the states are not in the same band, i.e. with $E-E' \gg \sigma$, then their contributions are small. Thus, we can use the following approximation for the $\delta$-function
\begin{equation}
    \delta(x)=\lim_{\sigma\rightarrow0}\frac{1}{\pi}\frac{\sigma x^2}{(x^2+\sigma^2)^2}\,.
\end{equation}
This gives us
\begin{equation}
    \begin{split}
         \overline{\bra{g,a}[\hat{H}_{\text{matter}},\hat{D}]\ket{f,a}}=&-i\sum_{k}\int dE dE'\rho(E) f_{k}(E)\rho(E')g_{k}(E')C(E,E')\delta(E-E')+\mathcal{O}(\frac{\sigma}{\Delta E})\\=&-i\sum_{k}\int dE\rho(E)\rho(E)C(E,E)f_{k}(E)g_{k}(E)+\mathcal{O}(\frac{\sigma}{\Delta E})\,,\label{eq:HDinstates}
    \end{split}
\end{equation}
where $\Delta E$ is the typical energy difference between states on neighbouring energy bands and it should be bigger than the typical width of each band. We will take $C(E,E)$ as
\begin{equation}
    C(E,E)=\frac{1}{\rho(E)}\,,
\end{equation}
which implies that
\begin{equation}
    \begin{split}
         \overline{\bra{g,a}[\hat{H}_{\text{matter}},\hat{D}]\ket{f,a}}=-i\sum_{k}\int dE\rho(E)f_{k}(E)g_{k}(E)+\mathcal{O}(\frac{\sigma}{\Delta E})\,,\label{eq:HDinstates}
    \end{split}
\end{equation}
The goal is to compare the result Equ.~(\ref{eq:HDinstates}) with the inner product
\begin{equation}
    \overline{\bra{g,a}f,a\rangle}=\overline{\sum_{\alpha,k,l}f_{k}(E_{\alpha})g_{l}(E_{\alpha})\sum_{n\in \mathcal{B}_{\alpha}}a^{k}_{n}a_{n}^{l*}}=\sum_{\alpha,k} f(E_{\alpha})g(E_{\alpha})\rho(E_{\alpha})\delta E_{\alpha}=\sum_{k}\int dE f_{k}(E)g_{k}(E)\,.
\end{equation}

As a result, we have
\begin{equation}
    \bra{f,a}[\hat{H}_{\text{matter}},\hat{D}]\ket{g,a}=-i\bra{f,a}\ket{g,a}+\mathcal{O}(\frac{\sigma}{\Delta E})\,.
\end{equation}
Let's denote the entropy within this energy band as $S$ i.e.
\begin{equation}
    \text{number of states within the band}= e^{S}\sim\frac{\Delta E}{\sigma }\,,
\end{equation}
where we used the fact that $\sigma$ is larger than the level spacing but their ratio doesn't scale with $e^{S}$.\footnote{So as $\Delta E$ versus the band width, i.e. $\Delta E$ is larger than the band width but their ratio doesn't scale with $e^{S}$.}

Hence we effectively have
\begin{equation}
    [\hat{H}_{\text{matter}},\hat{D}]=-i+\mathcal{O}(e^{-S})\,,\label{eq:clock}
\end{equation}
within the set of states $\ket{f,a}$.

Now we are ready to construct operators that approximately commute with the ADM Hamiltonian.

Note that the states under consideration here have a non-trivial spread in energy over a dense part of the spectrum. One example would be significant matter distribution in AdS, like a planet. Another could be a large black hole, in a pure state. Although such backgrounds could admit coarse grained isometries, we here allow ourselves to construct operators using the exact microstates, which resolve the states into highly time dependent ones. In the special case where the state appears time dependent even to coarse grained observables, the quantum clock construction reproduces a semi-classical clock. 

Given a bulk QFT operator dressed to the boundary, we can construct an operator that approximately commutes with the ADM Hamiltonian using the approximate clock operator $\hat{D}$. As an explicit example, if we are given a scalar QFT operator $\hat{\phi}(t)$ then we can construct
\begin{equation}
    \hat{O}^{\text{Phys}}(t)=\hat{\phi}(t-\hat{D})\,.\label{eq:sdo2}
\end{equation}
Using Equ.~(\ref{eq:clock}) and the basic quantum mechanical relation $[\hat{H}_{\text{ADM}},\hat{\phi}(t)]=[\hat{H}_{\text{matter}},\hat{\phi}(t)]=-i\partial_{t}\hat{\phi}(t)$, we can check
\begin{equation}
    [\hat{H}_{\text{ADM}},\hat{\phi}(t-\hat{D})]=0+\mathcal{O}(e^{-S})\,.
\end{equation}
As a result, the ability for a near-boundary observer to detect a bulk excitation created by
\begin{equation}
    \hat{U}=e^{i\alpha \hat{O}^{\text{Phys}}(t)}\,,
\end{equation}
using our protocol in Sec.~\ref{sec:no locality} is suppressed by a factor of order $e^{-S}$. A relevant example where our construction applies is a class of black hole microstates constructed out of the energy eigenstates in the energy window around the mass of the black hole following Equ.~(\ref{eq:statewithfeature}), for which the suppression factor is $e^{-S_{BH}}$ with $S_{BH}$ as the Bekenstein-Hawking entropy.

Moreover, the approximate relation Equ.~(\ref{eq:clock}) is in fact robust under the deformation of the states by simple operators like $\hat{\phi}$, i.e. operators which create only a few particles. This is because we can consider the $\hat{\phi}$-deformed states
\begin{equation}
    \ket{f,b}=f(\hat{H}_{\text{ADM}})\hat{\phi}\ket{a}=\sum_{n}b^{i}_{n}f_{i}(E_{n})\ket{E_{n}}\,,
\end{equation}
where we have a different set of microscopic parameters $b^{i}_{n}$ satisfying Equ.~(\ref{eq:ensemble}) and again slowly varying macroscopic wavefunctions $f$. In this case one can similarly prove that
\begin{equation}
    \bra{f,I}[\hat{H}_{\text{ADM}},\hat{D}]\ket{g,J}=i\bra{f,I}g,J\rangle+\mathcal{O}(\frac{\delta}{\Delta E})\,,
\end{equation}
where $I,J=a,b$ as long as we choose $R_{mn}$ such that
\begin{equation}
    R_{mn}^{ij}=C_{mn}a_{n}^{i*}a_{m}^{j}+C_{mn}b_{n}^{i*}b_{m}^{j}+\gamma_{n}^{i*}\gamma_{m}^{j}\,,
\end{equation}
where the vector $\vec{\gamma}$ is another random vector independent of $\vec{a}$ and $\vec{b}$. One can in fact extend this construction to include all deformed states by introducing more independent random vectors.

Finally, we want to remind the readers that these microstate dressed operators we just constructed exits for typical black hole microstates and so they are distinct from the operators we constructed in Sec.~\ref{sec:sdo} which have a simple perturbative bulk description. For example, the later operators don't exist for the old black holes.

\subsection{Dressing to Entanglement}\label{sec:dresstoENTG}
As opposed to Sec.~\ref{sec:approxlocality} and Sec.~\ref{sec:sdo}, the above construction is an inherently quantum construction where the clock operator $\hat{D}$ is constructed by specifying all its matrix elements inside the energy window we are considering. This motivates us to consider a more quantum construction by which the operator is dressed to quantum features of a single state. Such features are encoded in the microscopic entanglement structure of the state. Nevertheless, for our purpose it is enough to use the position dependent features. A natural feature of this type is described by correlations between operators inserted at different locations. Physically, this indicates that with a local source turned on in a spacetime, one is able to construct a coordinate system relative to this source using the quantum correlation between the source and other places as coordinates.

To be explicit, let's consider the case with an empty AdS$_{d+1}$ background in the Poincar\'{e} patch and all quantum fields have zero vacuum expectation values. Let's take two massive free scalar fields $\hat{\phi}_{\text{obj}}(x)$ and $\hat{\phi}_{\text{ref}}(x,z)$. We will construct a clock operator $\hat{D}$ from the reference field $\hat{\phi}_{\text{ref}}(x,z)$ and use this clock operator to dress the field $\hat{\phi}_{\text{obj}}(x,z)$. Let's denote the CFT dual of $\hat{\phi}_{\text{ref}}(x,z)$ as $\hat{O}_{\text{ref}}(y)$ with conformal dimension $\Delta$. We have the vacuum correlator
\begin{equation}
    \langle\hat{\phi}_{\text{ref}}(x,z)\hat{O}_{\text{ref}}(y)\rangle=\frac{z^{\Delta}}{[z^2+(\vec{x}-\vec{y})^{2}-(t_{x}-t_{y})^{2}]^\Delta}\,,\label{eq:quantumbackground}
\end{equation}
between the boundary and the bulk operators in the background AdS metric $g^{0}_{\mu\nu}$. 

Let's denote
\begin{equation}
  \hat{K}=\hat{\phi}_{\text{ref}}(x,z)\partial_{t_{y}}\hat{O}_{\text{ref}}(y)-\langle\hat{\phi}_{\text{ref}}(x,z)\partial_{t_{y}}\hat{O}_{\text{ref}}(y)\rangle\,,
\end{equation}
as the fluctuating part of the composite operator $\hat{\phi}_{\text{ref}}(x,z)\partial_{t_{y}}\hat{O}_{\text{ref}}(y)$ with the boundary reference point $y$ fixed. Under the bulk timelike diffeomorphism
\begin{equation}
    t\rightarrow t+\epsilon(x)\,,\label{eq:timetrans}
\end{equation}
we have
\begin{equation}
    \hat{K}\rightarrow \hat{K}+\epsilon(x) \partial_{t_{x}}\hat{\phi}_{\text{ref}}(x,z)\partial_{t_{y}}\hat{O}_{\text{ref}}(y)\,.
\end{equation}

To the leading order in perturbation theory, we have
\begin{equation}
    \hat{K}\rightarrow \hat{K}+\epsilon(x) \langle\partial_{t_{x}}\hat{\phi}_{\text{ref}}(x,z)\partial_{t_{y}}\hat{O}_{\text{ref}}(y)\rangle\,.
\end{equation}
Thus, we have a clock operator
\begin{equation}
    \hat{D}
    =-\frac{\hat{K}}{\langle\partial_{t_{x}}\hat{\phi}_{\text{ref}}(x,z)\partial_{t_{y}}\hat{O}_{\text{ref}}(y)\rangle}\,,
\end{equation}
which transforms under the timelike diffeomorphism Equ.~(\ref{eq:timetrans}) as
\begin{equation}
    \hat{D}\rightarrow\hat{D}-\epsilon(x)\,.
\end{equation}

As a result, we can define the dressed operator
\begin{equation}
    \hat{\Phi}(t)=\hat{\phi}(t+\hat{D})\,,
\end{equation}
where we ignored the dressing in the spacelike directions for simplicity. The above dressed operator is invariant under the small bulk diffeomorphisms, which act trivially on the boundary. Nevertheless, it doesn't commute with the bulk matter Hamiltonian $\hat{H}_{\text{matter}}$ as the boundary operator $\hat{O}(y)$ doesn't commute with it. More precisely, we have
\begin{equation}
    [\hat{H}_{\text{matter}},\hat{D}]=0\,,
\end{equation}
where
\begin{equation}
    \hat{H}_{\text{matter}}=\int d^{d}x\sqrt{g^{0}}N\hat{\mathcal{H}}_{\text{matter}}\,,
\end{equation}
which is the matter Hamiltonian density integrated over the full bulk Cauchy slice. Therefore, this dressing cannot be used to hide the bulk excitation created by $U=e^{i\Phi(t)}$ from the boundary observer, who is operating our protocol in Sec.~\ref{sec:no holography}.

We note that this construction equally applies in setups where ${\cal O}_{\text{ref}}$ is in an entangled, but external or even decoupled, system from the AdS spacetime, for example as in the island scenario in Fig.~\ref{pic:AdSbath}.

However, due to the fact that such a dressing used a bilocal operator, as opposed to line operators in the gravitational Wilson line construction in \cite{Donnelly:2015hta}, the excitation created by $\hat{U}=e^{i\hat{\Phi}(t)}$ can be hidden from an observer sitting slightly inside the bulk operating a similar protocol as in Sec.~\ref{sec:no locality}. Such an observer has access to the ADM Hamiltonian $\hat{H}_{\text{ADM}}^{\epsilon}$ Equ.~(\ref{eq:ADM}) evaluated at a cutoff surface $z=\epsilon$ close to the asymptotic boundary. Due to the constraint, $\hat{H}^{\epsilon}_{\text{ADM}}$ implements time translation for  operators living inside the bulk bounded by this cutoff surface. Thus, we have
\begin{equation}
    [\hat{H}^{\epsilon}_{\text{ADM}},\hat{O}(y)]=0\,,
\end{equation}
and so
\begin{equation}
    [\hat{H}^{\epsilon}_{\text{ADM}},\hat{D}]=i\,.
\end{equation}
As a result, we have
\begin{equation}
    [\hat{H}^{\epsilon}_{\text{ADM}},\hat{\Phi}(t)]=0\,,
\end{equation}
which implies that the excitation created by $\hat{U}=e^{i\hat{\Phi}(t)}$ is not detectable by such an observer using our protocol. 

We note that the construction in this subsection is an explicit example of the idea of \textit{dressing to entanglement}. This idea is proposed in \cite{Geng:2025rov}. The construction in this section and \cite{Geng:2025rov} are explicit examples of such dressing. However, we note that a recent paper \cite{Antonini:2025sur} contains some arguments in Section 4.7 against the validity of this dressing. The main point of \cite{Antonini:2025sur} was that the entanglement of only a few fields may not be strong enough  a feature to allow a reliable dressing of a nonzero number of bits. This is because \cite{Antonini:2025sur} speculated that, as opposed to a classical background field configuration, quantum entanglement may be too fluctuating and not be sufficiently localized to make use of in localizing a bulk point. The reasoning in \cite{Antonini:2025sur}  is that a classical background field configuration has a definite profile, so it is easily measured, but to measure quantum correlators like Equ.~(\ref{eq:quantumbackground}) we might need a large number of measurements.\footnote{We thank Henry Maxfield for clarifying this point.} We will show explicitly below why this is not correct.

First, the difficulty to measure the correlator Equ.~(\ref{eq:quantumbackground}) can be understood by comparing the correlator with the variance of it. The variance is given by
\begin{equation}
\begin{split}
    \sigma_{\phi O}^2&=\langle\big(\hat{\phi}_{\text{ref}}(x,z)\hat{O}_{\text{ref}}(y)-\langle\hat{\phi}_{\text{ref}}(x,z)\hat{O}_{\text{ref}}(y)\rangle\big)^2\rangle\,,\\&=\langle\hat{\phi}_{\text{ref}}(x,z)\hat{O}_{\text{ref}}(y)\hat{\phi}_{\text{ref}}(x,z)\hat{O}_{\text{ref}}(y)\rangle-\langle\hat{\phi}_{\text{ref}}(x,z)\hat{O}_{\text{ref}}(y)\rangle^2\,,\\&=\langle\hat{\phi}_{\text{ref}}(x,z)\hat{\phi}_{\text{ref}}(x,z)\rangle\langle\hat{O}_{\text{ref}}(y)\hat{O}_{\text{ref}}(y)\rangle+\langle\hat{\phi}_{\text{ref}}(x,z)\hat{O}_{\text{ref}}(y)\rangle^2\,.\label{eq:variancephiO}
    \end{split}
\end{equation}
Thus, we have
\begin{equation}
    \sigma_{\phi O}^2\geq\langle\hat{\phi}_{\text{ref}}(x,z)\hat{\phi}_{\text{ref}}(x,z)\rangle\langle\hat{O}_{\text{ref}}(y)\hat{O}_{\text{ref}}(y)\rangle=\infty\,,
\end{equation}
where the second step used the fact that the short-distance correlators of local operators are divergent. As a result, the variance $\sigma_{\phi O}$ is infinitely bigger than the correlator Equ.~(\ref{eq:quantumbackground}). One might naively conclude from here that it takes an infinite number of measurements to measure the correlator Equ.~(\ref{eq:quantumbackground}) to a reasonable level of accuracy. However, as we discussed in Sec.~\ref{sec:smearing}, this is the usual Bohr-Rosenfield effect and the correct observables for a quantum field theory should be smeared local operators. Intuitively, if one replaces $\hat{\phi}_{\text{ref}}(x,z)$ and $\hat{O}_{\text{ref}}(y)$ by Gaussian smeared operators $\hat{\phi}_{\text{ref,s}}(x,z)$ and $\hat{O}_{\text{ref,s}}(y)$ with the smearing length scale $\mu$, then the variance Equ.~(\ref{eq:variancephiO}) becomes
\begin{equation}
\begin{split}
    \sigma_{\phi O,s}^2&=\langle\hat{\phi}_{\text{ref,s}}(x,z)\hat{\phi}_{\text{ref,s}}(x,z)\rangle\langle\hat{O}_{\text{ref,s}}(y)\hat{O}_{\text{ref,s}}(y)\rangle+\langle\hat{\phi}_{\text{ref,s}}(x,z)\hat{O}_{\text{ref,s}}(y)\rangle^2\,,\\&\sim\frac{z^{d-1}}{\mu^{d-1+2\Delta}}+\langle\hat{\phi}_{\text{ref,s}}(x,z)\hat{O}_{\text{ref,s}}(y)\rangle^2\,,
    \end{split}
\end{equation}
where we used the fact that in the short-distance limit we have
\begin{equation}
    \begin{split}
\langle\hat{\phi}_{\text{ref}}(\tilde{x})\hat{\phi}_{_{\text{ref}}}(\tilde{x}')\rangle&\rightarrow \frac{z^{\frac{d-1}{2}}z'^{\frac{d-1}{2}}}{\big((z-z')^2-(t-t')^2+(\vec{x}-\vec{x}^{\text{ }\prime})^2\big)^{\frac{d-1}{2}}}\,,\\\langle\hat{O}_{\text{ref}}(y')\hat{O}_{\text{ref}}(y)\rangle&=\frac{1}{\big((\vec{y}-\vec{y}^{\text{ }\prime})^2-(t_{y}-t_{y}')^2\big)^{\Delta}}\,.\label{eq:shortdistancephiO}
    \end{split}
\end{equation}
Moreover, if there are $K$ distinct species of reference fields $\hat{\phi}_{\text{ref,s,I}}(x,z)$ with $I=1,2,\cdots,K$ then we can consider instead
\begin{equation}
    \phi O_{s}=\sum_{I=1}^{K}\hat{\phi}_{\text{ref,s,I}}(x,z)\hat{O}_{\text{ref,s,I}}(y)\,,
\end{equation}
for whom we have
\begin{equation}
    \frac{\sigma_{\phi O_s}}{\langle\phi O_s\rangle}=\frac{\sqrt{\frac{Kz^{d-1}}{\mu^{d-1+2\Delta}}+\sum_{I=1}^{K}\langle\hat{\phi}_{\text{ref,s},I}(x,z)\hat{O}_{\text{ref,s},I}(y)\rangle^2}}{\langle\sum_{I=1}^{K}\hat{\phi}_{\text{ref,s},I}(x,z)\hat{O}_{\text{ref,s},I}(y)\rangle}\sim\frac{1}{\sqrt{K}}\,,
\end{equation}
for large $K$. Thus, when the number of species $K$ is large, the variance can be easily tamed. In fact, this observation is rather suggestive, as even for a single field there are an infinite number of orthogonal modes in a local region. Therefore, the variance can be tamed even if the species number $K$ is one, as long as one properly distills a large number of modes from the reference field $\hat{\phi}_{\text{ref}}(x,z)$ and considers distinct modes as distinct species. In summary, to tame the variance Equ.~(\ref{eq:variancephiO}) with one reference field $\hat{\phi}_{\text{ref}}(x,z)$, one has to smear the field $\hat{\phi}_{\text{ref}}(x,z)$ and the dual boundary operator $\hat{O}_{\text{ref}}(y)$ and then properly distill a reasonably large number of orthogonal modes $\hat{\phi}_{\text{ref,s},I}(x,z)$ and $\hat{O}_{\text{ref,s},I}(y)$ from them.

\begin{figure}[h]
    \centering
    \includegraphics[width=0.5\linewidth]{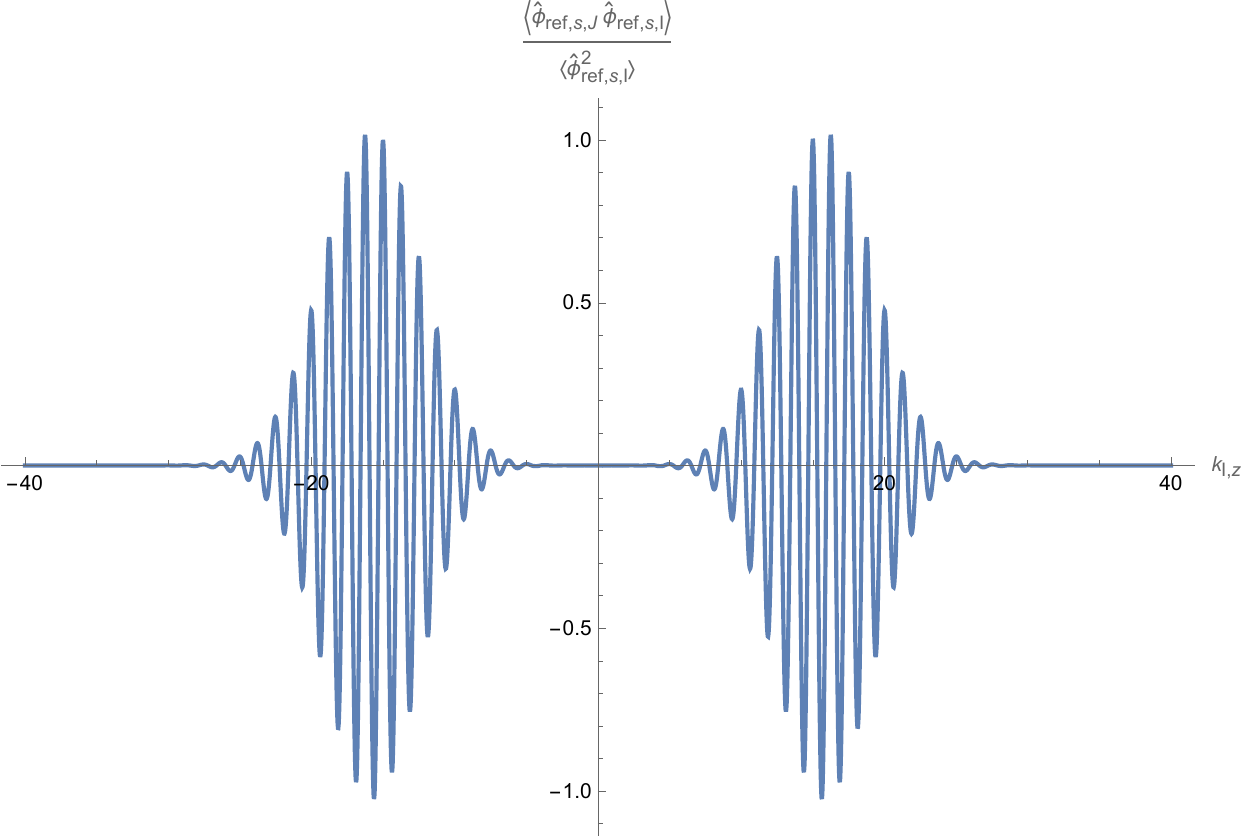}
    \caption{The ratio $\frac{\langle\hat{\phi}_{\text{ref,s},I}(x,z)\hat{\phi}_{\text{ref,s},J}(x,z)\rangle}{\langle\hat{\phi}_{\text{ref,s},I}(x,z)\hat{\phi}_{\text{ref,s},I}(x,z)\rangle}$ with $\mu=0.4, k_{J,z}=15$ and $\vec{x}=0,z=5$. From this plot we can see that the ratio is extremely small if $k_{I,z}$ is away from $\pm k_{J,z}$ enough.}
    \label{pic:phi}
\end{figure}

\begin{figure}[h]
    \centering
    \subfloat[ $k_{I,z}=\frac{2\pi}{5}(I+999)$]{
    \includegraphics[width=0.5\linewidth]{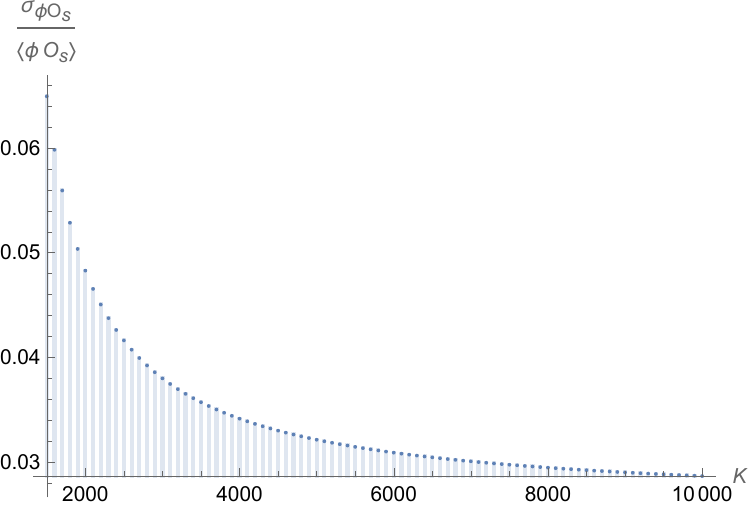}}
    \subfloat[$k_{I,z}=\frac{2\pi}{5}(I+4999)$]{
    \includegraphics[width=0.5\linewidth]{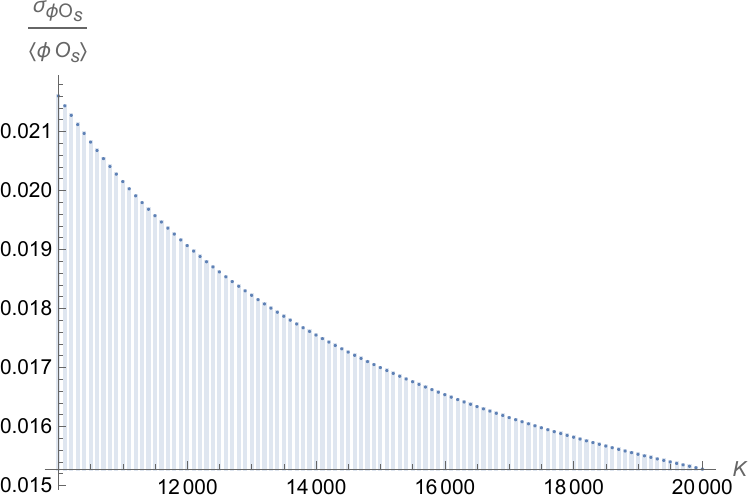}}
    \caption{The discrete plot of the ratio Equ.~(\ref{eq:rationfinal}) with $\mu=0.4, \vec{y}=0$ and $\vec{x}=0,z=5$ as a function of $K$, the number of distilled modes, for the choices $k_{I,z}=\frac{2\pi}{5}(I+999)$ and $k_{I,z}=\frac{2\pi}{5}(I+4999)$ with $I=1,2,\cdots,K$. From the plot we can see that the ratio is getting closer to zero as we distill more modes or with the first mode having a higher energy.}
    \label{pic:ratio}
\end{figure}

As a concrete example, let's consider a conformally coupled scalar field in AdS$_{4}$ quantized under the alternative quantization scheme. Such a scalar field has mass square $m^2=-\frac{d^2}{4}+1=-\frac{5}{4}$ and the conformal weight of the dual CFT operator is
\begin{equation}
    \Delta=\frac{d}{2}-\sqrt{\frac{d^2}{4}+m^2}=\frac{d-1}{2}=1\,.
\end{equation}
The bulk two-point function and the bulk-to-boundary correlator of this field are given by
\begin{equation}
    \begin{split}
        \langle\hat{\phi}_{\text{ref}}(\tilde{x})\hat{\phi}_{_{\text{ref}}}(\tilde{x}')\rangle&= \frac{z^{\frac{d-1}{2}}z'^{\frac{d-1}{2}}}{\big((z-z')^2-(t-t')^2+(\vec{x}-\vec{x}^{\text{ }\prime})^2\big)^{\frac{d-1}{2}}}+\frac{z^{\frac{d-1}{2}}z'^{\frac{d-1}{2}}}{\big((z+z')^2-(t-t')^2+(\vec{x}-\vec{x}^{\text{ }\prime})^2\big)^{\frac{d-1}{2}}}\,,\\\langle\hat{O}_{\text{ref}}(y')\hat{O}_{\text{ref}}(y)\rangle&=\frac{1}{\big((\vec{y}-\vec{y}^{\text{ }\prime})^2-(t_{y}-t_{y}')^2\big)^{\frac{d-1}{2}}}\,,
    \end{split}
\end{equation}
with $d=3$ and for which the bulk two-point function is the same as the flat space massless scalar field with a Neumann boundary at $z=0$ up to a conformal factor. This fact will significantly simplify our analysis. In fact, for our purpose, it is safe to ignore the second term in the bulk two-point function as it just indicates the boundary condition. We consider the following modes of this field $\hat{\phi}_{\text{ref}}(x,z)$ and the dual CFT operator $\hat{O}_{\text{ref}}(y)$
\begin{equation}
    \begin{split}
\hat{\phi}_{\text{ref,s},I}(x,z)&=\int dz'd^{2}\vec{x}'\frac{1}{z'(\sqrt{2\pi}\mu)^3}e^{-\frac{(z-z')^2+(\vec{x}-\vec{x}')^2}{2\mu^2}}\cos (k_{I,z}z+\vec{k}_{I}\cdot\vec{x})\hat{\phi}_{\text{ref}}(x',z')\,,\\\hat{O}_{\text{ref,s},I}(y)&=\int d^{2}\vec{y}'\frac{1}{(\sqrt{2\pi}\mu)^2}e^{-\frac{(\vec{y}-\vec{y}')^2}{2\mu^2}}\cos (\vec{k}_{I}\cdot\vec{y}')\hat{O}_{\text{ref}}(y')\,,\label{eq:distill}
    \end{split}
\end{equation}
where the smearing is only done in the spatial directions and to a good level of approximation the $z'$ integral can be extended from $(0,\infty)$ to $(-\infty,\infty)$ due to the Gaussian damping and $z>0$. Furthermore, in the first equation of Equ.~(\ref{eq:distill}), the factor $\frac{1}{z'}$ performs a Weyl transform of the to be smeared and distilled bulk field $\hat{\phi}_{\text{ref}}$ from AdS$_{4}$ to flat space. As a result, we are effectively studying the flat space bulk two-point function with a simple plane wave convoluted Gaussian smearing. For the sake of concreteness, let's consider $\mu=0.4,\vec{x}=0$ and $z=5$. We will consider $k_{I,z}$'s and their difference to be large enough such that
\begin{equation}
    \frac{\langle\hat{\phi}_{\text{ref,s},I}(x,z)\hat{\phi}_{\text{ref,s},J}(x,z)\rangle}{\langle\hat{\phi}_{\text{ref,s},I}(x,z)\hat{\phi}_{\text{ref,s},I}(x,z)\rangle}\,,\frac{\langle\hat{\phi}_{\text{ref,s},I}(x,z)\hat{O}_{\text{ref,s},J}(y)\rangle}{\langle\hat{\phi}_{\text{ref,s},I}(x,z)\hat{O}_{\text{ref,s},I}(y)\rangle}\,,\frac{\langle\hat{O}_{\text{ref,s},I}(y)\hat{O}_{\text{ref,s},J}(y)\rangle}{\langle\hat{O}_{\text{ref,s},I}(y)\hat{O}_{\text{ref,s},I}(y)\rangle}\ll1\,,\quad\text{for } I\neq J\,.
\end{equation}
This can be easily achieved if $k_{I,z}$ and the difference $k_{I,z}-k_{J,z}$ for $I\neq J$ are both large enough compared to $\frac{1}{\mu}$ (see Fig.~\ref{pic:phi} for a demonstration). Thus, we have
\begin{equation}
    \frac{\sigma_{\phi O_s}}{\langle\phi O_s\rangle}=\frac{\sqrt{\sum_{I=1}^{K}\langle\hat{\phi}_{\text{ref,s},I}(x,z)\hat{O}_{\text{ref,s},I}(y)\rangle^2+\sum_{I=1}^{K}\langle\hat{\phi}_{\text{ref,s},I}(x,z)^2\rangle\langle\hat{O}_{\text{ref,s},I}(y)^2\rangle}}{\sum_{I=1}^{K}\langle\hat{\phi}_{\text{ref,s},I}(x,z)\hat{O}_{\text{ref,s},I}(y)\rangle}\,,\label{eq:rationfinal}
\end{equation}
and this ratio can be much smaller than one if one properly chooses $k_{I,z}$. As a particular example, we take $k_{I,z}=\frac{2\pi}{5}(I+999)$ and $k_{I,z}=\frac{2\pi}{5}(I+4999)$ for $I=1,2,\cdots,K$ and as a function of the number $K$ of the distilled modes we have the plot for Equ.~(\ref{eq:rationfinal}) in Fig.~\ref{pic:ratio}. As a result, we can conclude that the variance Equ.~(\ref{eq:variancephiO}) can be tamed by our smearing and distilling protocol. 

\subsection{Island setup-- Massive Gravity and the Quantum Wormhole Operator}\label{sec:massive}
\begin{figure}[h]
    \centering
    \begin{tikzpicture}[scale=0.8]
       \draw[-,very thick] 
       decorate[decoration={zigzag,pre=lineto,pre length=5pt,post=lineto,post length=5pt}] {(-2.5,0) to (2.5,0)};
       \draw[-,very thick,red] (-2.5,0) to (-2.5,-5);
       \draw[-,very thick,red] (2.5,0) to (2.5,-5);
         \draw[-,very thick] 
       decorate[decoration={zigzag,pre=lineto,pre length=5pt,post=lineto,post length=5pt}] {(-2.5,-5) to (2.5,-5)};
       \draw[-,very thick] (-2.5,0) to (2.5,-5);
       \draw[-,very thick] (2.5,0) to (-2.5,-5);
       \draw[-,very thick,green] (-2.5,0) to (-5,-2.5);
       \draw[-,very thick,green] (-5,-2.5) to (-2.5,-5);
        \draw[-,very thick,green] (2.5,0) to (5,-2.5);
       \draw[-,very thick,green] (5,-2.5) to (2.5,-5);
       \draw[fill=green, draw=none, fill opacity = 0.1] (-2.5,0) to (-5,-2.5) to (-2.5,-5) to (-2.5,0);
       \draw[fill=green, draw=none, fill opacity = 0.1] (2.5,0) to (5,-2.5) to (2.5,-5) to (2.5,0);
       \draw[->,very thick,black] (-2.2,-3.5) to (-2.2,-1.5);
       \node at (-1.8,-2.5)
       {\textcolor{black}{$t$}};
        \draw[->,very thick,black] (2.2,-3.5) to (2.2,-1.5);
       \node at (1.8,-2.5)
       {\textcolor{black}{$t$}};
       \draw[-,thick,blue] (5,-2.5) arc (60:75.5:10);
       \draw[-,thick,blue] (-5,-2.5) arc (120:104.5:10);
       \draw[-,thick,blue] (2.5,-1.47) to (0,-2.5);
       \draw[-,thick,blue] (-2.5,-1.47) to (0,-2.5);
       \draw[-,thick,red] (5,-2.5) arc (60:70:10);
       \draw[-,thick,red] (-5,-2.5) arc (120:110:10);
        \draw[fill=red, draw=none, fill opacity = 0.2] (-5,-2.5) to (-5+1.2,-2.5+1.2) to (-3.42020143326,-1.76332782999) to (-5+0.4,-2.5-0.4) to (-5,-2.5);
         \draw[fill=red, draw=none, fill opacity = 0.2] (5,-2.5) to (5-1.2,-2.5+1.2) to (3.42020143326,-1.76332782999) to (5-0.4,-2.5-0.4) to (5,-2.5);
       \node at (-3.8,-2.2)
       {\textcolor{red}{$R_{I}$}};
        \node at (3.8,-2.2)
       {\textcolor{red}{$R_{II}$}};
       \draw[-,very thick, purple] 
       (0,-2.5) to (1.25,-1.985);
        \draw[-,very thick, purple] (0,-2.5) to (-1.25,-1.985);
 \draw[fill=purple, draw=none, fill opacity = 0.2]  (-1.25,-1.985) to (0,-1.985+1.25) to  (1.25,-1.985) to (0,-1.985-1.25) to  (-1.25,-1.985);
        \node at (0,-2.1)
       {\textcolor{purple}{$\mathcal{I}$}};
    \end{tikzpicture}
    \caption{We show the Penrose diagram of an eternal black hole in AdS$_{d}$ coupled to $d$-dimensional baths. The bath is the green shaded region. This is a scenario in the so called island setup. We choose the time evolution as indicated in the diagram. We also specify one Cauchy slices of this time evolution, where an island emerges, as the blue curve. We denote the domain of dependence $D(R)$ of the subsystem $R=R_{I}\cup R_{II}$ in red and the domain of dependence of the island $D(I)$ in purple. We emphasize that $D(I)$ overlaps with the black hole interior. }
    \label{pic:AdSbath}
\end{figure}
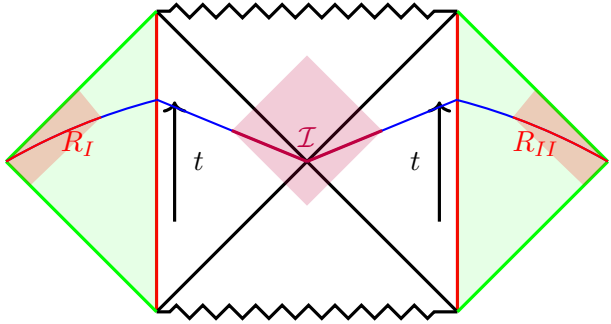

Based on the observation that graviton is massive in the island setup \cite{Almheiri:2019psf,Penington:2019npb}, it was suggested that the tension between gravitational constraints and locality can be relaxed if the graviton is massive \cite{Geng:2021hlu}. This was based on the intuition that in gauge theory if the gauge boson is massive then the potential of the force mediated by the gauge boson obeys a Yukawa form which is hence short-ranged. Thus, information about the bulk charge can be hidden from being detected using Gauss' law and its implications near the asymptotic infinity. Detailed underlying mechanism of this intuition is worked out in \cite{Geng:2023zhq,Geng:2025rov,Geng:2025byh}. The essential point is that in the island setup we have a gravitational AdS coupled with a non-gravitational bath (see Fig.~\ref{pic:AdSbath}) and this bath coupling prepares an entangled state between the AdS and the bath. This particular entanglement structure spontaneously breaks the diffeomorphism symmetry in the gravitational AdS. Hence, the AdS graviton becomes massive due to the Higgs mechanism. Here we revisit the arguments in \cite{Geng:2021hlu,Geng:2023zhq} for the sake of completeness of a collection of all known situations where our particle detecting protocol fails. 

The holographic interpretation of entanglement island is that the physics inside the island $\mathcal{I}$ is fully encoded in a bath subregion $R$ which claims the island (see Fig.~\ref{pic:AdSbath}). Nevertheless, such an interpretation is in tension with the fact that in a standard gravitational universe the Hamiltonian is a boundary term localized near the asymptotic boundary of this gravitational universe. In fact, this tension can be clearly formulated if one considers the setup with an empty AdS$_{d+1}$ coupled with a bath (i.e. replacing the black hole of Fig.~\ref{pic:AdSbath} by an empty AdS) and all the matter fields are taken to be in the ground state. Entanglement island also exists in this setup. In this setup, one can take an operator $\hat{O}(x)$ inside the island $\mathcal{I}$. Then according to our calculations in Sec.~\ref{sec:reviewH}, if the gravitational theory is the standard gravitational theory then we would have
\begin{equation}
    [\hat{O}(x),\hat{H}_{\partial}]\neq0\,.
\end{equation}
This result is in tension with the above holographic interpretation of the island. This is because in the bath $\hat{O}(x)$ is supposed to be dual to an operator localized inside $R$, while $\hat{H}_{\partial}$ is an operator that is localized in its complement $\bar{R}$. Thus, the operator $\hat{O}(x)$ inside the island should commute with $\hat{H}_{\partial}$.

As it was pointed out in \cite{Geng:2021hlu}, this trouble is avoided because the graviton is in fact massive in the island setup. We start from the fact that the graviton is massive in the island setup. The essential observation in \cite{Geng:2021hlu} is that when the graviton is massive the gravitational constraint Equ.~(\ref{eq:core}) is modified to
\begin{equation}
\mathcal{H}^{(1)}=-\frac{1}{\sqrt{16\pi G_{N}}}\Big[(d-1)h+\nabla_{i}\nabla_{j}h^{ij}-\nabla^{2}h+m^2 h\Big]+\mathcal{H}_{\text{matter}}=0\,,\label{eq:coremassive}
\end{equation}
where $m^2$ is the mass square of the graviton. Hence now the constraint Equ.~(\ref{eq:commutator}) for bulk operators is modified to
\begin{equation}
    [\sqrt{16\pi G_{N}}\hat{H}_{\text{matter}}-\hat{H}_{\partial}-m^2\int d^{d}x \sqrt{g_{0}}N\hat{h},\hat{O}(x)]=0\,,\label{eq:commutatormassive}
\end{equation}
which tells us that it is no longer true that we must have $[\hat{H}_{\partial},\hat{O}(x)]\neq 0$. 

Nevertheless, an explicit solution of the constraint Equ.~(\ref{eq:commutatormassive}) which commutes with $\hat{H}_{\partial}$ is constructed only recently in \cite{Geng:2023zhq}. The insight of \cite{Geng:2023zhq} is that one should view massive gravity as describing the Higgs phase of an underlying manifestly diffeomorphism invariant theory in the unitary gauge. To construct an explicit solution of Equ.~(\ref{eq:commutatormassive}), one should restore the fully diffeomorphism invariant description which requires a Goldstone vector $V^{\mu}(x)$ that transforms nonlinearly under diffeomorphism as
\begin{equation}
    V^{\mu}(x)\rightarrow V^{\mu}(x)-\epsilon^{\mu}(x)\,,\quad\text{while } x^{\mu}\rightarrow x^{\mu}+\sqrt{16\pi G_{N}}\epsilon^{\mu}(x)\,.
\end{equation}
Moreover, this vector field is divergenceless. The constraint in the fully diffeomorphism invariant description is
\begin{equation}
    [\sqrt{16\pi G_{N}}\hat{H}_{\text{matter}}-\hat{H}_{\partial}-2m^2\int d^{d}x \sqrt{g_{0}}N(\hat{h}+2\nabla_{i}\hat{V}^{i}),\hat{O}(x)]=0\,,\label{eq:commutatormassiveV}
\end{equation}
where we note that the combination $h_{\mu\nu}+\nabla_{\mu}V_{\nu}+\nabla_{\nu}V_{\mu}$ is diffeomorphism invariant \cite{Geng:2023ynk,Geng:2023zhq} and the conjugate momentum of $V^{\mu}(x)$ is
\begin{equation}
    \hat{\pi}^{\mu}_{V}=-2m^{2}\int d^{d}x\sqrt{g_{0}}N (h_{0}^{\mu}+\nabla_{0}V^{\mu}+\nabla^{\mu}V_{0})\,,
\end{equation}
whose zeroth component, due to the tracelessness of the graviton field $h_{\mu\nu}$ and divergenceless of the vector field $V^{\mu}(x)$, equals to
\begin{equation}
    \hat{\pi}^{0}_{V}=2m^{2}\int d^{d}x\sqrt{g_{0}}N (\hat{h}+2\nabla_{i}\hat{V}^{i})\,.
\end{equation}
Hence, the constraint Equ.~(\ref{eq:commutatormassiveV}) becomes
\begin{equation}
    [\sqrt{16\pi G_{N}}\hat{H}_{\text{matter}}-\hat{H}_{\partial}-\hat{\pi}^{0}_{V},\hat{O}(x)]=0\,.\label{eq:commutatormassiveVV}
\end{equation}
Therefore one can easily write a solution
\begin{equation}
    \hat{O}(x)=\hat{\phi}(x+\sqrt{16\pi G_{N}}\hat{V})\,,\label{eq:solution}
\end{equation}
which only depends on the diffeomorphism invariant combination $x^{\mu}+\sqrt{16\pi G_{N}}\hat{V}^{\mu}(x)$ and here $\hat{\phi}(x)$ is a scalar matter field. Other tensorial solutions to Equ.~(\ref{eq:commutatormassiveVV}) can be easily constructed in a similar manner. One can explicitly check that
\begin{equation}
     [\sqrt{16\pi G_{N}}\hat{H}_{\text{matter}}-\hat{\pi}^{0}_{V},\hat{O}(x)]=0\,,
\end{equation}
using the fact that $\hat{\pi}^{0}_{V}$ is the conjugate momentum of $\hat{V}^{0}(x)$. Hence we have
\begin{equation}
[\hat{H}_{\partial},\hat{O}(x)]=0\,.
\end{equation}
Interestingly, one can go to the unitary gauge, i.e. fixing the diffeomorphism gauge symmetry by setting $V^{\mu}(x)=0$, in which $\hat{O}(x)$ becomes $\hat{\phi}(x)$ which looks to be a standard local field in the bulk.

We want to emphasize that the construction of the operator Equ.~(\ref{eq:solution}) in island setup that commutes with $\hat{H}_{\text{ADM}}$ should also be understood as a type of dressing to the feature of state which though is more universal and well-defined. This is due to that fact that in the Higgs phase the diffeomorphism is really spontaneously broken which is a feature of the ground state and the vector Goldstone field is just parametrizing the moduli space of gauge equivalent vacua \cite{Geng:2023zhq}. Nevertheless, as noticed in \cite{Geng:2020qvw} that in the case of the island setup the gravitational theory is fundamentally modified in the sense that its holographic dual is no longer a CFT$_{d}$ living on the asymptotic boundary of AdS$_{d+1}$. Instead, an additional bath is introduced which is coupled to the AdS$_{d+1}$ at the asymptotic boundary exchanging energy with the AdS$_{d+1}$ bulk \cite{Geng:2023ynk}. Therefore, our particle detecting protocol should still work if one uses the total Hamiltonian with the bath included instead of $\hat{H}_{\text{ADM}}$.\footnote{More precisely, if one takes the bath to be a half-space CFT$_{d+1}$ then the holographic dual of the setup with AdS$_{d+1}$ coupled to this bath is a boundary conformal field theory (BCFT$_{d+1}$). Our protocol will work if one uses the total Hamiltonian $\hat{H}_{\text{tot}}$ of this dual BCFT$_{d+1}$ instead of $\hat{H}_{\text{ADM}}$. The symmetry generated by $\hat{H}_{\text{tot}}$ is an isometry of the whole system so any local operator $\hat{O}(x)$ will have a commutator with $\hat{H}_{\text{tot}}$ as $[\hat{H}_{\text{tot}},\hat{O}(x)]=-i\partial_{t}\hat{O}(x)$.} Furthermore, the nature of the Goldstone vector field $V^{\mu}(x)$ is recently uncovered in \cite{Geng:2025rov} and it is in fact nonlocally supported on both the bath and the gravitational AdS$_{d+1}$. This fact explains why the dressed operator Equ.~(\ref{eq:solution}) will excite the bath but not the boundary of the AdS$_{d+1}$ and this is the mechanism behind the seemingly nonlocal information encoding scheme for island. More interestingly, the Goldstone vector field $V^{\mu}(x)$ can in fact be interpreted as a \textit{quantum wormhole} operator that connects the island and the bath and this interpretation is nicely geometrized in the Karch-Randall braneworld model \cite{Karch:2000gx,Karch:2000ct} where an extra-dimension exists connecting the island and the bath \cite{Geng:2025rov}. In fact, from this perspective, the clock operator we constructed in Sec.~\ref{sec:dresstoENTG} can also be interpreted as a quantum wormhole operator. From the algebraic perspective, this above analysis suggests the following relationships between the algebra of operators at the level of gravitational perturbation theory:
\begin{equation}
    \quad\mathcal{A}_{R}^{\text{Non-Pert}}\supseteq A_{\mathcal{I}\cup R}^{\text{Pert}}\supsetneq\mathcal{A}_{R}^{\text{Pert}}\,,
\end{equation}
where $
\mathcal{A}^{\text{Pert}}_{(.)}$ denotes the algebra of operators that is strictly localized inside $(.)$ at the level of gravitational perturbation theory. Moreover, from the above analysis we know that there is no nontrivial operators strictly localized inside the island $\mathcal{I}$. Thus, we also have
\begin{equation}
    \mathcal{A}_{\text{island}}^{\text{Pert}}=\mathbb{C}\,,\label{eq:strongconclusion}
\end{equation}
which only consists of $c$-numbers.

Since the result Equ.~(\ref{eq:strongconclusion}) looks very strong and one might wonder how general it is, let us make a few remarks about Equ.~(\ref{eq:strongconclusion}) before we wrap up this section. One might think that in the case of an evaporating black one could dress operators inside the island to the clock defined by the black hole evaporation. These operators are essentially the type of operators we considered in Sec.~\ref{sec:sdo} and they in principle also commute with the ADM Hamiltonian to all orders in gravitational perturbation theory. However, there are a few issues with such operators if one interprets them as physical observables that fundamentally act only in the bath Hilbert space:
\begin{enumerate}
    \item One can consider black holes that are evaporating slowly. In these cases, there are still islands emergent at the late-time of the evaporation. Then operators dressed to the internal features of the island will be delocalized in this scenario.
    \item One can consider a bath with an integrable discrete energy spectrum coupled with a black hole. In this case, one still has a Page curve and an island at late times if the energy gap of the bath is less than or equal to the energy of a single Hawking particle, i.e. $\mathcal{O}(T_{\text{black hole}})$. This energy gap is perturbatively visible, and so one expects that operators in the island will visibly excite the bath. However, this is not satisfied by operators dressed to the black hole evaporation.\footnote{This is satisfied instead by the operators Equ.~(\ref{eq:solution}) \cite{Geng:2025rov}.}  Therefore, there seems to be a tension if such black hole evaporation dressed operators are reconstructable by fine-grained operators that act strictly in the bath, i.e. belonging to $\mathcal{A}_{R}^{\text{Non-Pert}}$.
    \item They are not dressed to the essential feature of the states that enables the islands, i.e. the matter field entanglement between the islands and the regions that claim the islands.
\end{enumerate}
Thus, we conclude that even though at the level of the gravitational perturbation theory there could be other operators that are localized inside the islands which naively invalidate Equ.~(\ref{eq:strongconclusion}) in generic island setups, it deserves to be better understood if these operators are consistent with the holographic interpretation of entanglement islands.

\section{The Dual CFT Perspective on the Bulk Information Localizability}\label{sec:CFT}
In this section, we translate our understanding of bulk information localizablity in the previous sections from the bulk to the boundary CFT. We will first study the case corresponding to the bulk empty AdS$_{d+1}$ with all fields in the ground state. This is the case we studied in Sec.~\ref{sec:no locality} and we will provide a CFT perspective on the fact that bulk operators cannot be dressed to commute with the boundary ADM Hamiltonian $H_{\text{ADM}}$. Then we will study the case that there are some features in the bulk as that in Sec.~\ref{sec:sdo2}. We will provide a CFT interpretation for why this case differentiates from the previous case that bulk operators could be dressed to features to commute with the ADM Hamiltonian up to good approximations. For the sake of simplicity we will consider the Poincar\'{e} patch for the bulk AdS$_{d+1}$
\begin{equation}
    ds^{2}=\frac{-dt^{2}+d\vec{x}_{d-1}^{2}+dz^{2}}{z^{2}}\,,
\end{equation}
for which the dual CFT is living on a Minkowski background near the asymptotic boundary $z\rightarrow0$. We will follow the discussions in \cite{Kabat:2012av}.

\subsection{The Case with no Localizable Information in the Bulk}
\label{sec:CFT1}
The basic correlator we are studying is
\begin{equation}
    \langle [\hat{H},\hat{\phi}(x,z)]\hat{O}(y)\rangle\,,\label{eq:bulkcommutator}
\end{equation}
where $\hat{H}$ is the ADM Hamiltonian which is dual to the CFT Hamiltonian, $\hat{\phi}(x,z)$ is a bulk field and $\hat{O}(y)$ is its dual CFT primary operator. The nonvanishingness of the above correlator can be used to detect a bulk particle-like excitation that was created by the unitary operator Equ.~(\ref{eq:unitary}) following our protocol in Sec.~\ref{sec:no locality}. The above nonvanishingness is a result of the nonvanishingness of the following correlator
\begin{equation}
     \langle [\hat{H},\hat{O}(x')]\hat{O}(y)\rangle\,,\label{eq:commutatorO}
\end{equation}
as the bulk field $\phi(x,z)$ and the dual boundary operator $\hat{O}(x)$ are related as
\begin{equation}
    \hat{\phi}(x,z)=\int d^{d}x' K^{\text{HKLL}}_{\Delta}(x,z|x')\hat{O}(x')\,,\label{eq:phiO}
\end{equation}
with $K^{\text{HKLL}}_{\Delta}(x,z|x')$ is the HKLL kernel \cite{Hamilton:2006az} and the integration domain is over the boundary point $x'$ that is spacelike separated from the bulk point $(x,z)$ \cite{Hamilton:2006fh}. The correlator Equ.~(\ref{eq:commutatorO}) is nonzero due to the singularity of the following correlator as $w\rightarrow x'$
\begin{equation}
    \langle T_{0}^{0}(w)\hat{O}(x')\hat{O}(y)\rangle\,,\label{eq:correlator}
\end{equation}
where $T^{0}_{0}(w)$ is the $00$-component of the CFT stress-energy tensor whose spatial integral is $\hat{H}$. Due to the conformal symmetry, the ground state correlator Equ.~(\ref{eq:correlator}) has the universal form
\begin{equation}
    \langle T_{0}^{0}(w)\hat{O}(x')\hat{O}(y)\rangle=f_{OOT}\frac{Z^{0}Z_{0}-\frac{1}{d}Z^{\mu}Z_{\mu}}{|w-x'|^{d-2}|w-y|^{d-2}|x'-y|^{2\Delta-d+2}}\,,
\end{equation}
where we have
\begin{equation}
    Z^{\mu}=\frac{(w-x')^\mu}{|w-x'|^{2}}-\frac{(w-y)^{\mu}}{|w-y|^{2}}\,.
\end{equation}
Thus, as $w\rightarrow x'$ we have
\begin{equation}
    \langle T_{0}^{0}(w)\hat{O}(x')\hat{O}(y)\rangle\rightarrow f_{OOT}\frac{(w-x')^{0}(w-x')_{0}-\frac{1}{d}|w-x'|^{2}}{|w-x'|^{d+2}|x'-y|^{2\Delta}}\,.\label{eq:target}
\end{equation}

Nevertheless, Equ.~(\ref{eq:phiO}) is the bulk field operator in the limit that all interactions are turned off including gravity. The effects of interactions are captured by adding more operators on the left hand side of Equ.~(\ref{eq:phiO}), i.e. we have
\begin{equation}
    \hat{\phi}(x,z)=\int d^{d}x' K^{\text{HKLL}}_{\Delta}(x,z|x')\hat{O}(x')+\sum_{i}a_{i}\int d^{d}x' K^{\text{HKLL}}_{\Delta_{i}}(x,z|x')\hat{O}_{i}(x')\,,\label{eq:intphiO}
\end{equation}
where the operators $\hat{O}_{i}(x)$ are in general multi-trace primary operators with conformal weights $\Delta_{i}$ \cite{Kabat:2011rz,Kabat:2012av}. As we discussed in Sec.~\ref{sec:no holography}, when gravity is dynamical in the bulk AdS$_{d+1}$ we have to dress bulk operators to obey the diffeomorphism constraints. This is an effect of gravitational interaction and it is also captured by Equ.~(\ref{eq:intphiO}). Thus, we can use Equ.~(\ref{eq:intphiO}) to translate the question of whether we can have localized bulk excitations that are not detectable using our protocol in Sec.~\ref{sec:no locality} to the question of if we can add other operators to the dual boundary operator $\hat{O}(x)$ such that the singularity in Sec.~\ref{eq:target} is canceled. We note that the strength of the singularity in fact doesn't affect the corresponding bulk correlator Equ.~(\ref{eq:bulkcommutator}) \cite{Kabat:2012av}. Thus, the important thing is in fact the $y$-dependent coefficient of the singularity, i.e. $\frac{1}{|x-y|^{2\Delta}}$.

In this section, we study the CFT dual of the scenario in Sec.~\ref{sec:no locality}, where the bulk is empty AdS$_{d+1}$ with all fields in the ground state, so the dual CFT is also in the ground state. Thus all operators we have are CFT multi-trace primary operators and we don't have any nontrivial state-dependent operators like that we constructed in Sec.~\ref{sec:sdo2}. Moreover, the CFT stress-energy tensor $T_{\mu\nu}(x)$ is conserved so we cannot use operators like $\nabla_{\mu}T^{\mu\nu}(x)$ to construct multi-trace primary operators. The coefficient $\frac{1}{|x'-y|^{2\Delta}}$ implies that we should consider adding multi-trace primary operators of the form
\begin{equation}
    \hat{O}_{i}(x')=\hat{K}(x')\hat{O}(x')\,,\label{eq:Oi}
\end{equation}
where $\hat{K}(x')$ is some scalar CFT operator such that $\hat{O}_{i}(x')$ is primary. The singular behavior as $w\rightarrow x'$ for the corresponding correlator of this operator is
\begin{equation}
     \langle T_{0}^{0}(w)\hat{K}(x')\hat{O}(x')\hat{O}(y)\rangle\rightarrow\langle T^{0}_{0}(w)\hat{K}(x')\rangle\langle\hat{O}(x')\hat{O}(y)\rangle=\frac{\langle T^{0}_{0}(w)\hat{K}(x')\rangle}{|x'-y|^{2\Delta}}\,.
\end{equation}
If there exists a scalar operator $\hat{K}(x)$ such that $\langle T^{0}_{0}(w)\hat{K}(x')\rangle$ is nonzero then Equ.~(\ref{eq:Oi}) can be used to correct the boundary dual of the dressed bulk operator such that the correlator Equ.~(\ref{eq:bulkcommutator}) for dressed bulk operator $\hat{\phi}^{\text{dressed}}(x)$ is zero, which means that there can be bulk excitations that are not detectable by our protocol in Sec.~\ref{sec:no locality}. Nevertheless, in the current case with the CFT in the ground state, the only candidate operator for $\hat{K}(x')$ is $\nabla_{\mu}\nabla_{\nu}T^{\mu\nu}(x)$ which is however zero. As a result, in this case we are considering there is no operator $\hat{O}_{i}(x')$ that can be used to cancel the singularity in Equ.~\ref{eq:target}, i.e. there is no way to dress the bulk field $\hat{\phi}(x,z)$ such that it can be used to created excitations that are not detectable by the near-boundary observer using our protocol in Sec.~\ref{sec:no locality} \cite{Kabat:2012av}. This is consistent with our result in Sec.~\ref{sec:no locality}.

\subsection{The Case with Localizable Information in the Bulk}\label{eq:CFT2}
Now let's consider the cases where we have bulk state dressed operators as in Sec.~\ref{sec:sdo}, Sec.~\ref{sec:sdo2} and Sec.~\ref{sec:massive}. As we discussed earlier, these operators can be used to create bulk excitations that are not detectable by the near-boundary observer using our protocol in Sec.~\ref{sec:no locality}. Unlike the case we considered in Sect.~\ref{sec:CFT1}, in these cases we should have scalar operators $\hat{K}(x)$ such that $\langle T^{0}_{0}(x)\hat{K}(x')\rangle$ is nonzero. For example, in the case of Sec.~\ref{sec:massive} the CFT is coupled to a bath so $\nabla_{\mu}\nabla_{\nu} T^{\mu\nu}(x)$ is not zero and we can use it as the operator $K(x)$. The situation in Sec.~\ref{sec:sdo2} works similarly as we can take the operator $K(x)$ such that a proper spatial integral of it is $e^{i\hat{D}}$. Therefore, to demonstrate that $\langle T^{0}_{0}(w)\hat{K}(x')\rangle$ is singular as $w\rightarrow x'$, it is enough to check that $\langle [\hat{H},e^{i\hat{D}}]\rangle$ is nonzero. This quantity is nonzero as we have
\begin{equation}
\langle[\hat{H},e^{i\hat{D}}]\rangle=\langle e^{i\hat{D}}\rangle=1\,.
\end{equation}
The case in Sec.~\ref{sec:sdo} follows the same consideration as the emergent clock operator $\hat{D}$ can be easily constructed following the recipe in \cite{Geng:2024dbl}. We leave the question of a detailed CFT presentation of the operator $\hat{K}(x')$ in situations like Sec.~\ref{sec:sdo} and Sec.~\ref{sec:sdo2} to future work.

\section{Relation to Previous Works and Future Questions}\label{sec:relation}
In this section, we comment on the relationship between our work and previous works in the literature and point out questions to be studied in various directions.
\subsection{Particle Detection in AdS/CFT}
Particle detection in AdS/CFT was firstly studied in the early days of the development of this duality in \cite{Balasubramanian:1999zv} where a concrete protocol to detect a bulk particle in a global AdS$_{3}$ was provided. The essential observation in \cite{Balasubramanian:1999zv} is that a particle in the bulk of AdS$_{3}$ whose mass is below the black hole threshold \cite{Banados:1992wn} sources a conical deficit in the bulk which is a global orbifolding of the empty AdS$_{3}$. One can show that the correlators of the CFT operators  whose bulk dual is a probe quantum field with heavy mass will have kinks. This can be seen easily using the WKB method in the bulk, where the dominant geodesic computing the dual CFT correlator can jump due to the orbifolding. These kinks in the CFT correlators are proposed to be a good boundary probe of the bulk particle in \cite{Balasubramanian:1999zv}.\footnote{See \cite{Caron-Huot:2022lff} for similar ideas.} Nevertheless, it is unclear how this protocol could be generalized to higher dimensions and for more general bulk local excitations. In comparison, the protocol proposed in our paper could be used to detect a large class of bulk excitations described by Equ.~(\ref{eq:psi}) in empty AdS$_{d+1}$ for any $d\geq2$ (the $1+1$-d Einstein's gravity is trivial although there is a suitable generalization of our protocol  in Jackiw-Teitelboim gravity \cite{Teitelboim:1983ux,Jackiw:1984je,Grumiller:2002nm,Maldacena:2016upp,Mertens:2022irh} as it also satisfies the gravitational Gauss' law \cite{Geng:2023zhq}).\footnote{In JT gravity, the analogy to higher dimensional vacuum AdS in global patch is to consider a negative tension end-of-the-world brane. Otherwise one wouldn't be able to have a gapped ground state \cite{Gao:2021uro}.} Though the particle detection in the situation discussed in Sec.~\ref{sec:sdo} should be better understood.

\subsection{The Tension between Locality and Holography}\label{sec:giddingsexp}
Holography and locality are generally in tension with each other. The former states that the physics in a $(d+1)$-dimensional system is fully encoded in a local $d-$dimensional system which is clearly in tension with the locality in $(d+1)$-dimension. The usual statement reconciling these two statements is that the breakdown of locality wouldn't manifest until the Planck scale is reached where the Compton wavelength of a particle would reach its Schwarzschild radius and hence forms a black hole. This statement is later generalized to scattering processes in \cite{Giddings:2001pt} and \cite{Ghosh:2016fvm,Ghosh:2017pel} where the former states that in the two-particle scattering process the impact parameter should be large enough otherwise the two scattered particles with center-of-mass energy $E$ will form a black hole before the scattering process happens and the later states that the perturbative expansion of gravitational scattering wouldn't give a convergent result if the number of particles involved in a scattering process is large enough. In this former statement, it is important to note that particle scattering is a manifestation of locality as it is captured by a local quantum field theory and the later work show that a perturbative calculation of such scattering process involving gravity will break down if a large number of particles are involved in the scattering process. In a sense the later statement and the former statement are equivalent to each other as when a large number of particles are involved in a scattering process the impact parameter for individual scattering is necessarily suppressed. These considerations were trying to defer the breakdown of locality in quantum gravity to the small length scale regime where nonperturbative effects like the Planck scale black hole formation and evaporation dominate. Nevertheless, the perturbative breakdown of locality at large distance scale due to the gravitational constraints in certain background was later noticed in \cite{Giddings:2006vu,Giddings:2012gc,Giddings:2013kcj,Giddings:2013noa,Donnelly:2015hta,Donnelly:2017jcd,Giddings:2017mym,Donnelly:2018nbv,Giddings:2019hjc} which however was argued to be not enough to be used to extract any bulk information. The argument for this later point was that measuring the boundary Poincar\'{e} charges including the Hamiltonian wouldn't be enough to distinguish bulk configurations with equal Poincar\'{e} charges. However, it was notice in \cite{Hsu:2013fra,Raju:2018zpn,Chowdhury:2020hse} that correlators between the boundary Hamiltonian and other boundary operators are in principle enough to fully reconstruct any bulk state in the empty AdS$_{d+1}$ background. Our paper provides an explicit such reconstruction method for a large class of bulk states motivated by \cite{Chowdhury:2020hse} and demonstrated that this reconstruction can be done efficiently in a short time, i.e. the data processing time is small if the bulk excitation is sufficiently localized in time Equ.~(\ref{eq:dp}). A recent objection of such type of protocols by \cite{Giddings:2021khn} is that the correlation between the boundary Hamiltonian and the boundary operators which is the core of such reconstruction protocols is exponentially suppressed in flat space. In this paper, we noticed the AdS version of this suppression is due to the decay of the bulk field $\phi(\rho,t,\Omega)$ as $\rho$ approaches its boundary value. This is a standard behavior of bulk field operators in AdS/CFT and our claim is that we should think of the near-boundary observer as the dual CFT observer for whom the actual operators to use are the extrapolated versions of the bulk operators following Equ.~(\ref{eq:O}). Hence our near-boundary observer shouldn't be understood as a QFT observer. As a result, we see that the perturbative breakdown of locality in empty AdS spacetime gives a version of holography \cite{Chowdhury:2021nxw} which is practically relevant to efficiently extract bulk information from the boundary. 

In fact, in the context of black holes \cite{Hsu:2008yi,Hsu:2013cw,Hsu:2013fra,Hsu:2020jbc,Calmet:2021stu,Calmet:2022bpo,Calmet:2023met,Calmet:2023gbw,Calmet:2024tgm} followed upon \cite{Page:1979tc} argued that one is able to reconstruct the full pure state of the black hole and its radiation from such perturbative gravitational backreactions.\footnote{See also \cite{Nomura:2012cx,Nomura:2012ex,Nomura:2012sw,Nomura:2013gna,Nomura:2018kia,Nomura:2019dlz,Nomura:2019qps,Nomura:2020ewg,Nomura:2020ska,Concepcion:2024nwv} for relevant works.} However, we note that in a black hole geometry with a perturbatively sharp horizon, the spectrum of the ADM Hamiltonian is continuous, so one naturally expects that some information can be hidden in such a spacetime from being perturbatively detectable from the asymptotic boundary.\footnote{As the commutator between the ADM Hamiltonian and an operator describing the perturbative fluctuations can be extremely small due to the continuous spectrum of $H_{\text{ADM}}$.} Moreover, we believe that the large distance perturbative breakdown of bulk locality in AdS should be encoded in the structures of the dual boundary CFT. A relevant work is \cite{Chen:2017dnl} which we believe will be useful to understand this question.

\subsection{Dressing to Observer and the Issue of Internal Time}\label{sec:observer}
It is recently argued in \cite{Chandrasekaran:2022cip,Witten:2023qsv,Witten:2023xze,Jensen:2023yxy,AliAhmad:2023etg,DeVuyst:2024pop,Faulkner:2024gst,Franzmann:2024rzj} that a neat construction of bulk operators that satisfy the Hamiltonian constraint Equ.~(\ref{eq:co}) and meanwhile commute with the boundary ADM Hamiltonian can be done in the presence of a bulk observer which is modeling an experimentalist whose Hamiltonian $\hat{H}_{\text{obs}}$ appears together with $\hat{H}_{\text{matter}}$ in the Hamiltonian constraint. The intuition is that bulk operators can be dressed to this observer and it is claimed that such a construction can be done for generic background geometries and matter distributions. 

A similar mechanism can be understood in a simple example - an eternal AdS$_{d+1}$ black hole, where the left boundary plays the role of the observer. In this case the bulk has two exterior regions $L$ and $R$ which are separated by the black hole bifurcation point and local operators supported in region $R$ can satisfy the Hamiltonian constraint by dressing it to the boundary of the region $L$. More precisely, in this case the integrated version of the Hamiltonian constraint on bulk operators is
\begin{equation}
[\sqrt{16\pi G_{N}}\hat{H}_{\text{matter}}+\hat{H}_{\text{$\partial$,L}}-\hat{H}_{\text{$\partial$,R}},\hat{O}(x)]=0\,.
\end{equation}
Hence for an operator $\hat{O}_{\text{R}}(x)$ in the region $R$ we can choose the following construction
\begin{equation}
[\hat{H}_{\text{$\partial$,L}},\hat{O}_{\text{R}}(x)]=-\sqrt{16\pi G_{N}}[\hat{H}_{\text{matter}},\hat{O}_{\text{R}}(x)]=-i\sqrt{16\pi G_{N}}\partial_{t}\hat{O}_{\text{R}}(x)\,,
\end{equation}
where $t$ is the exterior time in the eternal black hole geometry which conjugates to the exterior timelike Killing vector. In this way we would have 
\begin{equation}
[\hat{H}_{\text{$\partial$,R}},\hat{O}_{\text{R}}(x)]=0\,,
\end{equation}
and the explicit dressing using the gravitational Wilson line going from $x$ to the boundary of the $L$ region can be written down using techniques developed in \cite{Donnelly:2015hta}. This is related to the crossed-product construction of \cite{Witten:2021unn}. We should nevertheless notice that this construction doesn't give an operator that is strictly localized in the region $R$ due to the Wilson line dressing. The interesting feature of the bulk operator constructed in this way is that it cannot be detected by a near-boundary observer in region $R$ using our protocol in Sec.~(\ref{sec:no locality}) as it commutes with the Hamiltonian in the $R$ boundary. It is instead an operator that can only be detected with access to the two boundaries i.e. using $\hat{H}_{\text{ADM,R}}-\hat{H}_{\text{ADM,L}}$ and other boundary operators in the $R$ boundary. This is not surprising as the bulk geometry is connected and so bulk information is in principle encoded in the two boundaries. 

This analysis motivates the idea that we might be able to define a 
subregion algebra for a more generic subregion $R$ in a gravitational spacetime $\Sigma_{g}$ that satisfies the Hamiltonian constraint and commutes with the ADM Hamiltonian on the boundary $\partial\Sigma_{g}$ by putting an observer inside $R$, such that $R$ is the causal diamond of the observer, and dressing all operators to this observer \cite{Chandrasekaran:2022cip,Witten:2023qsv,Witten:2023xze,Jensen:2023yxy,AliAhmad:2023etg}. More specifically, the Hamiltonian constraint is satisfied and the resulting operator commutes with the ADM Hamiltonian on the boundary if we have
\begin{equation}
[\hat{H}_{\text{matter}}+\hat{H}_{\text{obs}},\hat{O}(x)]=0\,.\label{eq:obscons}
\end{equation}
Of course, this then requires a specific modelling of the observer whose Hamiltonian should have a canonical conjugate that is not the coordinate time but a nontrivial operator $\hat{\tau}$ which can be thought of as the internal time or an emergent clock. Then a solution of Equ.~(\ref{eq:obscons}) can be easily written as
\begin{equation}
\hat{O}(x)=e^{-i\hat{\tau}\hat{H}_{\text{matter}}}\hat{\phi}(x)e^{i\hat{\tau}\hat{H}_{\text{matter}}}\,,
\end{equation}
where $\hat{\phi}(x)$ is the bulk matter field. An important observation from the above black hole analysis that motivates such a model of an observer is that the ADM Hamiltonian is first order in the metric fluctuation or graviton field $\hat{h}_{\mu\nu}$ so its canonical conjugate in the bulk can be easily constructed using the canonical conjugate of $\hat{h}_{\mu\nu}$. This construction can be explicitly done in simple spacetimes \cite{Donnelly:2015hta}. Hence \cite{Chandrasekaran:2022cip,Witten:2023qsv,Witten:2023xze,Jensen:2023yxy} models the observer by a Hamiltonian
\begin{equation}
\hat{H}_{\text{obs}}=\hat{p}\,,
\end{equation}
where the operator $\hat{p}$ has a conjugate operator $\hat{q}$ such that
\begin{equation}
[\hat{p},\hat{q}]=-i\,.
\end{equation}
For a physical observer, the operator $\hat{p}$ will be a complicated operator that depends on the details of the observer, and $\hat{q}$ will be an emergent clock.

Nevertheless, it is unclear what is the precise physical principle that determines the emergence of such an operator and also it is unclear if the operator $\hat{p}$ itself has to be properly dressed to the boundary in the case of a spacetime with a spatial boundary for example the empty AdS$_{d+1}$. Our construction in Sec.~\ref{sec:sdo2} provides a primary study of this question. We found that we can have approximate clock operators emergent from backgrounds with strong enough features. Though it is still unclear if there could exist exact clock operators as proposed by \cite{Chandrasekaran:2022cip,Witten:2023qsv,Witten:2023xze,Jensen:2023yxy,AliAhmad:2023etg,DeVuyst:2024pop,Faulkner:2024gst,Franzmann:2024rzj}. Interestingly, we notice that a possible way to put the proposal in \cite{Chandrasekaran:2022cip,Witten:2023qsv,Witten:2023xze,Jensen:2023yxy,AliAhmad:2023etg,DeVuyst:2024pop,Faulkner:2024gst,Franzmann:2024rzj} into a concrete context is as follows. When we are trying to define a subregion algebra for a generic subregion $R$ we can cut open the full spacetime along its boundary $\partial R$ and impose the Dirichlet boundary condition for graviton field near $\partial R$. Then the integrated Hamiltonian constraint would be the same as Equ.~(\ref{eq:ham}) and it contains a boundary term $\hat{H}_{\partial}$ which is now localized on $\partial R$. Hence we can define algebra in this new region by dressing operators to $\partial R$ the same way as what we did in Sec.~\ref{sec:no locality}. Nevertheless, this construction introduces a boundary to the subregion $R$ instead of a simple observer inside it and it is unclear if such an algebra can be thought of as a subregion algebra in the original spacetime $\Sigma_{g}$.\footnote{Though for certain subregion $R$, for example whose boundary $\partial R$ is a (quantum) extremal surface, this might be possible \cite{Geng:2020fxl,Geng:2023qwm,Balasubramanian:2023dpj}.} Moreover, a recent work \cite{Geng:2025bcb} pointed out that one should think of the vector Goldstone boson in the island setup we discussed in Sec.~\ref{sec:massive} as an example of such an exact clock operator.

\subsection{Dressing to State and Features}\label{sec:dresstostate}
As we have discussed in Sec.~\ref{sec:no holography}, a different way to define local operators in a gravitational bulk is to dress it to features of the state, i.e. the background configuration. A recent construction in this spirit is provided in \cite{Bahiru:2022oas,Bahiru:2023zlc} in the AdS/CFT context, which is however not explicitly done in the bulk. Our constructions in Sec.~\ref{sec:sdo} and Sec.~\ref{sec:sdo2} provide a direct bulk construction of such operators. Our construction suggests that we need strong enough features of the state to define generic local operators in the bulk. For example, simple features like a uniform dust of particles, an end of world brane (next to which the gravity is not turned off, see Sec.~\ref{sec:observer}), a uniform matter shell or a spherical shockwave couldn't be used to define local bulk operators. The reason is that operators dressed to them are not localized in all directions since the backgrounds have unfixed isometries.

This intuition is also supported by the results in \cite{Bahiru:2022oas,Bahiru:2023zlc} which involve time-smeared projected bulk operators in a highly excited state $\langle H\rangle\sim G_{N}^{-\frac{1}{d-1}}$ with a large energy variance $\delta E\sim G_{N}^{-\frac{1}{2(d-1)}}$. The projection is to the initial GNS Hilbert space and the time smearing is over a time scale of order $\mathcal{O}(1)$ in $G_{N}^{-\frac{1}{2(d-1)}}$ expansion.\footnote{That is, one needs the width of the smearing time band  $t_{*}$ to satisfy $\frac{1}{\delta E}\sim G_{N}^{\frac{1}{2(d-1)}}\ll t_{*}\ll G_{N}^{-\frac{1}{2(d-1})}$.} In this case, the highly excited state with a large energy variance can be treated as a classicalized background. The projection together with the time-smearing can be thought of as a diffusive operation which quickly diffuses or relaxes any local quantum excitations into this classicalized background.  
Their result is consistent with our result in Sec.~\ref{sec:sdo2} that the commutator between the ADM Hamiltonian and the relaxed operators is suppressed by the background entropy as $\mathcal{O}(e^{-S})$ and in the black hole context $S\sim \frac{1}{G_{N}}\sim N^{2}$ which reproduces the results in \cite{Bahiru:2022oas,Bahiru:2023zlc}. The fact that their construction involves microstate projectors makes it similar to our construction of the emergent clock operator in Equ.~(\ref{eq:clockoperator}). This should be contrasted to the constructions in other sections in this paper which only involve bulk macroscopic features (these operators exist in bulk perturbation theory). Moreover, our emergent clock is an improvement of their construction as the operators can be localized to very short time and it is limited only by the level spacing of the energy spectrum.

\subsection{A New Type of Islands Scenario}\label{sec:Island}
\begin{figure}[h]
\begin{centering}
\begin{tikzpicture}
\draw[-,very thick,red] (2,0) arc (0:360:2);
 \draw[fill=violet, draw=none, fill opacity = 0.2] (2,0) arc (0:360:2);
 \draw[-,very thick, blue] (0.5,0.5) arc (90:450:1 and 0.5);
 \draw[-,very thick, blue] (-1.414,1.414) arc (90:-90:0.7 and 1.414);
  \draw[fill=orange, draw=none, fill opacity = 0.5] (-1.414,1.414) arc (90:-90:0.7 and 1.414) arc (225:135:2);
 \draw[fill=orange, draw=none, fill opacity = 0.5] (0.5,0.5) arc (90:450:1 and 0.5);
 \node at (-0.2,0)   {\textcolor{black}{$\cross$}};
\end{tikzpicture}
\caption{\small{A special type of island as discussed in \cite{Geng:2021hlu} that could potentially exist when graviton is massless. We will call it a pseudo island. We draw a constant time slice in an asymptotic AdS geometry. Fields inside the bulk obeys the reflective boundary conditions near the asymptotic boundary (the red circle). The dual CFT is in a pure state. The entanglement wedge of a boundary subregion could potentially have two disconnected components, with one component containing this boundary subregion and another component purely reside inside the bulk (the two orange regions in figure). The component residing insider the bulk is the pseudo island but not an entanglement island. The cross represent an operator insertion inside the pseudo island.} }\label{pic:specialisland}
\end{centering}
\end{figure}
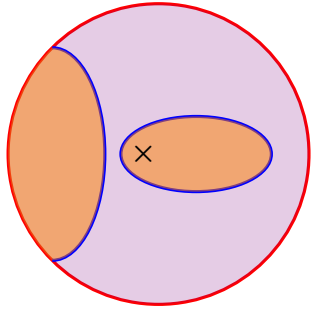

A possible island-like scenario that is distinct from entanglement islands in the context of black hole information paradox is shown in Fig.~\ref{pic:specialisland}, although there is not yet a completely explicit construction. Here the island region is a disconnected component of the entanglement wedge of a subregion of the holographic CFT, rather than of an external system, hence we will refer to that disconnected component of the entanglement wedge in this scenario as a \textit{pseudo island}. 

A potential puzzle in such constructions is that operators inside the pseudo islands cannot be dressed using gravitational Wilson lines as they would go outside the pseudo island. This naively suggests that pseudo islands could only exist where the bulk has additional features such as a nontrivial background matter profile as we discussed in Sec.~\ref{sec:sdo} \cite{Folkestad:2023cze}. These features significantly complicate the application of the quantum extremal surface formula to search for pseudo islands. However, an interesting implication of our construction in Sec.~\ref{sec:dresstoENTG} is that operators inside pseudo islands do not necessarily have to be dressed to local features. The entanglement between the bulk and the boundary in the semiclassical background can be used to dress operators inside the pseudo island to the pseudo island claiming boundary subregion without ever touching the complementary bulk region to the entanglement wedge of this boundary subregion (i.e. the purple part of Fig.~\ref{pic:specialisland}). Thus, pseudo islands could potentially exist in much simpler scenarios. We leave an explicit construction of this type of islands to future work but we should mention that a closely related scenario has been explicitly constructed in Section 3 of \cite{Geng:2020fxl}.

\subsection{Approximate Notions of Bulk Locality}\label{sec:approxloc}
In Sec.~\ref{sec:sdo2} and Sec.~\ref{sec:approxlocality} we constructed situations where we could have an approximate notion of locality in a gravitational bulk even for simple backgrounds whose geometries are pure AdS$_{d+1}$. These constructions are potentially useful in modeling an observer in a gravitational spacetime and the discussion of his/her experience. We leave the study of these questions to future work.

\section{Conclusions}\label{sec:con}
In this paper, we critically examined the fate of locality in a gravitational spacetime directly in the bulk. We considered asymptotically AdS$_{d+1}$ spacetimes with the AdS/CFT correspondence in our mind. We find that information localizability and the perturbative notion of holography are generally in tension with each other. This conclusion comes from our construction of situations where the perturbative notion of holography manifests and information localizability breaks down, and also situations where the information localizability persists and the perturbative notion of holography is obscure. Our tool is the Hamiltonian constraint, which is the basic constraint we have to impose if we want to study quantum gravity directly in a gravitational bulk. Our work has intimate connections with and implications to many works in the literature in various directions in the study of quantum gravity and holography. Many relevant questions are proposed to be studied in the near future to push our understanding of quantum gravity forward.

\acknowledgments
We are grateful to Stefano Antonini, Tom Hartman, Steve Giddings, Stephen Hsu, Yikun Jiang, Andreas Karch, Juan Maldacena, Henry Maxfield, Joseph Minahan, Yasunori Nomura, Geoff Penington, Massimo Porrati, Suvrat Raju, Lisa Randall, Martin Sasieta and Jiuci Xu for helpful discussions. We would like to thank Steve Giddings, Andreas Karch, Suvrat Raju and Barbara \v{S}oda for comments on the draft. 
HG would like to thank the hospitality from the Aspen Center for Physics where the final stage of this work is performed. Research at the Aspen Center for Physics is supported by National Science Foundation grant PHY-2210452 and a grant from the Simons Foundation (1161654, Troyer).  HG is supported by the Gravity, Spacetime, and Particle Physics (GRASP) Initiative from Harvard University and a grant from the Physics Department at Harvard University. The work of DLJ, PS and NT are supported in part by the Simons Investigator in Physics Award MP-SIP-0001737 and U.S. Department of Energy grant DE-SC0007870. 
\appendix

\bibliographystyle{JHEP}

\bibliography{references}

\end{document}